\documentclass{article}

\pdfoutput=1
\usepackage{arxiv}

\usepackage[utf8]{inputenc} 
\usepackage[T1]{fontenc}    
\usepackage[hidelinks]{hyperref}       
\usepackage{url}            
\usepackage{booktabs}       
\usepackage{amsfonts}       
\usepackage{nicefrac}       
\usepackage{microtype}      
\usepackage{lipsum}
\usepackage{graphicx}
\graphicspath{ {./images/} }
\usepackage{natbib}
\usepackage{amsmath}
\usepackage{makecell}
\usepackage{adjustbox}
\usepackage[table]{xcolor}
\usepackage{enumitem}

\usepackage{multirow}
\usepackage[title]{appendix}

\def\bSig\mathbf{\Sigma}

\usepackage{natbib}
\usepackage{amsmath}
\usepackage{mathtools}
\usepackage{makecell}
\usepackage{adjustbox}
\usepackage[table]{xcolor}
\usepackage{siunitx}
\usepackage[mathscr]{euscript}
\usepackage{amsfonts}
\usepackage{xcolor}
\usepackage{xr}
\usepackage[flushleft]{threeparttable}

\title{Exposure Effects on Count Outcomes with Observational Data, with Application to Incarcerated Women}

\author{\small{Bonnie E. Shook-Sa$^{1*}$, 
Michael G. Hudgens$^{1}$, Andrea K. Knittel$^{2}$, Andrew Edmonds$^{3}$, }\\ \textbf{\small{Catalina Ramirez$^{2}$, Stephen R. Cole$^{3}$, Mardge Cohen$^{4}$, Adebola Adedimeji$^{5}$, Tonya Taylor$^{6}$, }} \\ \textbf{\small{Katherine G. Michel$^{7}$, Andrea Kovacs$^{8}$, Jennifer Cohen$^{9}$, Jessica Donohue$^{10}$, Antonina Foster$^{11}$, }} \\ \textbf{\small{Margaret A. Fischl$^{12}$, Dustin Long$^{13}$, and Adaora A. Adimora$^{2,3}$}} \\ 
\footnotesize{$^{1}$Department of Biostatistics, University of North Carolina at Chapel Hill, Chapel Hill, North Carolina, U.S.A. $^{2}$School of } \\ \footnotesize{Medicine,University of North Carolina at Chapel Hill, U.S.A. $^{3}$Department of Epidemiology, University of North Carolina at }\\\footnotesize{Chapel Hill, U.S.A. $^{4}$Stroger Hospital, U.S.A. $^{5}$Albert Einstein College of Medicine, U.S.A. $^{6}$SUNY Downstate Medical Center, } \\ \footnotesize{  U.S.A. $^{7}$Department of Infectious Diseases,  Georgetown University, U.S.A. $^{8}$Keck School of Medicine, University of } \\ \footnotesize{Southern California, U.S.A. $^{9}$Department of Medicine, University of California, San Francisco, U.S.A. $^{10}$Department of} \\ \footnotesize{  Epidemiology, Johns Hopkins Bloomberg School of Public Health, U.S.A.  $^{11}$Department of Medicine, Emory University, U.S.A. } \\ \footnotesize{$^{12}$Division of Infectious Diseases, University of Miami Miller School Medicine, U.S.A. $^{13}$ The University of } \\ \footnotesize{Alabama at Birmingham, U.S.A.} \\ \footnotesize{$*$}\footnotesize{bshooksa@email.unc.edu}
}

\begin{document}
	
	\maketitle
	
	\begin{abstract}{
Causal inference methods can be applied to estimate the effect of a point exposure or treatment on an outcome of interest using data from observational studies. For example, in the Women's Interagency HIV Study, it is of interest to understand the effects of incarceration on the number of sexual partners and the number of cigarettes smoked after incarceration. In settings like this where the outcome is a count, the estimand is often the causal mean ratio, i.e., the ratio of the counterfactual mean count under exposure to the counterfactual mean count under no exposure. This paper considers estimators of the causal mean ratio based on inverse probability of treatment weights, the parametric g-formula, and doubly robust estimation, each of which can account for overdispersion, zero-inflation, and heaping in the measured outcome. Methods are compared in simulations and are applied to data from the Women's Interagency HIV Study.}
	\end{abstract}

	
\section{Introduction}
\label{sec:counts-intro}
Researchers often seek to estimate the causal effect of a point exposure or treatment on an outcome of interest. Randomized experiments are infeasible for many exposures, and thus inference often relies on data from observational studies. Associations measured from such studies can be subject to confounding, so various methods have been developed to consistently estimate causal effects from observational data. Three commonly-used methods are inverse probability of treatment weight (IPTW) estimators, \citep{robins1998marginal,Robins2000,hernan2000marginal}, the parametric g-formula \citep{robins1986new}, and doubly robust estimators that incorporate both exposure and outcome model estimators \citep{bang2005doubly,Hernan2010,FunkDR,kang2007demystifying}. In practice, these estimators are frequently applied to observational data when the outcome of interest is continuous, binary, or categorical \citep{hernan2000marginal, bodnar2004marginal,Cole2008, Taubman, young2011comparative, Garcia,FunkDR, waernbaum2012model}. Count outcomes are also common in observational studies, as researchers often seek to estimate measures over a fixed period of time such as numbers of sexual partners \citep{wiederman1997truth}, pill counts to assess treatment adherence \citep{bangsberg2001comparing}, or the number of cigarettes smoked \citep{singh1994smoothed}. For example, in this paper we aim to estimate the effects of incarceration on the number of sexual partners and the number of cigarettes smoked per day in the subsequent six-month period among women with HIV or at risk of acquiring HIV.

Incarceration of women has been rapidly increasing in recent decades, growing at twice the rate of men’s incarceration \citep{kajstura2019women}. Incarceration disproportionately affects persons with HIV and those at risk of acquiring HIV \citep{harawa2008incarceration}. Quantifying the effects of incarceration on health outcomes for vulnerable populations helps inform public health policy and practice. For example, incarceration can disrupt established relationships and alter women's sexual networks upon release from prison or jail. Because condom use has been shown to be inconsistent in some at-risk populations, increases in sexual partners can result in potential for transmission of sexually transmitted infections \citep{knittel}. Thus it is important to assess the effect of incarceration on subsequent sexual behavior.

Incarceration also has the potential to alter smoking behavior in women. Incarcerated women smoke at much higher rates than women in the general public \citep{Binswangerg4542}.  While smoking bans in prisons and jails have increased dramatically since the 1980s, they are not universal, implementation has been inconsistent, and obstacles to smoking cessation in incarcerated populations remain \citep{kennedy2014smoke,zhang2018prison}. Smoking relapse rates after release from incarceration are high \citep{zhang2018prison}. While the effect of incarceration on smoking behavior have been studied for some populations \citep{bailey2015incarceration}, no studies have estimated this effect for women with HIV or at risk of acquiring HIV. Adults with HIV are nearly twice as likely to smoke as adults without HIV \citep{mdodo2015cigarette}, and they are more likely to develop the serious health consequences of smoking (e.g., heart disease, cancer, infections) compared to adults without HIV \citep{hivgov}. For these reasons, it is critical to understand how incarceration affects smoking behavior in this vulnerable population.

In this paper, the effects of incarceration on the numbers of sexual partners and cigarettes smoked are estimated using data from the Women's Interagency HIV Study (WIHS), a multicenter cohort study of women with HIV or at risk of acquiring HIV \citep{adimora2018cohort}. At each six-month visit, the WIHS collects data regarding women's self-reported incarceration status, sexual behaviors, and substance use behaviors during the prior six-month period. To allow for estimation of a point exposure effect using this longitudinal data, the study sample includes all women from the WIHS who were incarcerated between 2007-2017 and a stratified random sample of women who remained unincarcerated throughout the study period. To ensure the same distribution across WIHS visits for incarcerated and unincarcerated participants, random sampling of unincarcerated women was stratified by WIHS visit.

Estimating the effect of incarceration on counts of sexual partners and cigarettes smoked poses challenges that must be accounted for in the analysis. The Poisson distribution is commonly used to model count outcomes, but the observed variance of a count outcome often exceeds the variance assumed under the Poisson model, i.e., there may be overdispersion. Zero-inflation occurs when the number of observed zero counts exceeds the number expected under the Poisson distribution \citep{bohning1999zero}. The number of sexual partners reported by WIHS participants exhibits both zero-inflation and overdispersion, while the number of reported cigarettes among smoking participants exhibits overdispersion. Count outcomes are also susceptible to data heaping, a form of measurement error which occurs when reported counts are rounded to different levels of precision \citep{wang2008modeling}. This phenomenon is commonly observed when collecting self-reported retrospective counts or measures of duration, including cigarette usage \citep{klesges1995self}, duration of breastfeeding \citep{singh1994smoothed}, and number of sexual partners \citep{wiederman1997truth,roberts2001measures}. For example, reported cigarette counts in the WIHS are heaped at multiples of ten. Data heaping is often attributed to cognitive processes in respondents, including choosing round numbers or approximations (digit preference) or using estimation methods to aid in recall \citep{roberts2001measures}. Often data heaping is informative in the sense that the probability of reporting an exact count versus rounding depends on the (unobserved) true outcome. For example, individuals with larger cigarette counts might be more likely to round their reported count compared to individuals with smaller counts \citep{klesges1995self}. Data heaping distorts the true underlying distributions of counts, which makes point and variance estimators that ignore this measurement error biased when applied to the observed data \citep{wang2008modeling}.

To estimate the effect of a binary point exposure on a count outcome, the estimand is often the causal mean ratio, which contrasts the counterfactual mean count under exposure to the counterfactual mean count under no exposure over a fixed period of time. Previous research has considered parametric g-formula estimators of the causal mean ratio for zero-inflated count outcomes \citep{albert2014estimating} and g-formula and Targeted Maximum Likelihood Estimation (TMLE) estimators for count outcomes in the longitudinal setting \citep{schnitzer2014effect}. While existing causal methods for count outcomes can yield valid inference in the presence of overdispersion and zero-inflation, bias may occur in the presence of data heaping. Furthermore, while several IPTW and g-formula approaches have been proposed to account for outcome measurement error outside of the heaping setting, there are seemingly few doubly robust estimators that accommodate outcome measurement error  \citep{shu2019causal}. In this paper we develop IPTW, parametric g-formula, and doubly robust estimators for the causal mean ratio, each of which can account for overdispersion, zero-inflation, and data heaping. 

The remainder of this paper is organized as follows. Section \ref{sec:counts-methods} presents the estimators in detail and describes their large sample properties. Section \ref{sec:counts-sim} demonstrates and compares the empirical properties of the estimators with a simulation study, and analyses of the WIHS data are presented in Section \ref{sec:counts-WIHS}. Section \ref{sec:counts-discussion} concludes with a discussion of the results. The Appendix includes proofs of the results appearing in the main text, supplemental tables and figures from the simulation study, and sensitivity analyses for the applications presented in Section \ref{sec:counts-WIHS}. R code for computing the different estimators along with the corresponding standard error estimators is available on GitHub. 

\section{Methods}
\label{sec:counts-methods} 

\subsection{Preliminaries}
\label{subsec:prelim}
Consider an observational study where the aim is to assess the effect of a binary exposure (or treatment) $A$ on an outcome $Y \in \mathbb{N}^0$, where $\mathbb{N}^0$ is the set of non-negative integers. In the data analysis in Section \ref{sec:counts-WIHS}, $A$ represents a woman's incarceration status and two outcomes $Y$ are considered, the number of male sexual partners and the number of cigarettes smoked per day, each measured in the subsequent six-month period. Let $L$ denote a vector of baseline covariates. For example, in the WIHS analysis $L$ includes a woman's age, race, drug use status, and additional covariates. Unless noted otherwise, all vectors are assumed to be row vectors. Assume $n$ independent and identically distributed copies of $(A,Y,L)$ are observed, denoted $(A_i,Y_i,L_i)$ for $i=1,...,n$. Let $Y_i^1$ denote the potential outcome if individual $i$, possibly counter to fact, is exposed. Similarly, let $Y_i^0$ denote the potential outcome if individual $i$ is not exposed, such that $Y_i=A_iY_i^1+(1-A_i)Y_i^0$. Assume that conditional exchangeability holds, i.e., $Y^a \perp A \mid L$, $a \in \{0,1\}$. Also assume that positivity holds such that $\Pr(A=a \mid L=l)>0$ for all $l$ such that $dF_L(l)>0$ and $a \in \{0,1\}$, where $F_L$ is the cumulative distribution function of $L$. Let $E(Y^a)=\lambda^a$ for $a \in \{0,1 \}$. The goal is to draw inference about the causal mean ratio, $CMR=\lambda^1/\lambda^0$. 
\subsection{Estimators: Correctly measured outcome}
\label{sec:methodsnoheap}
This section presents three estimators of the $CMR$ that are consistent and asymptotically normal when the outcome is measured without error.

\subsubsection{Inverse Probability of Treatment Weighting}
\label{sec:MSMmethods}
Consider the (saturated) marginal structural model (MSM) \begin{equation} \label{MSM} \log(\lambda^a) = \beta_0 + \beta_1 a
\end{equation} Under the assumptions specified in Section \ref{subsec:prelim}, the parameters of (\ref{MSM}), and hence $CMR=\exp(\beta_1)$, can be consistently estimated using IPTW as follows. First, the propensity score for each participant, $e_i=\Pr(A_i=1 \mid L_i)$, is estimated using a finite dimensional parametric model. For example, $A$ can be regressed on $L$ using logistic regression, i.e., the model $\mbox{logit}(e_i)=X_{i}\alpha$ is fit, where $X_i=g_p(L_i)$ is a vector of predictors for participant $i$ for some user-specified function $g_p$ of $L_i$ and $\alpha$ is the column vector of regression coefficients. Predicted propensity scores are calculated as $\hat{e}_i=e(L_i,\hat{\alpha})=\mbox{logit}^{-1}(X_i \hat{\alpha})$ where $\hat{\alpha}$ is the maximum likelihood estimate (MLE) of $\alpha$. Participant $i$'s IPTW is estimated as $\hat{W}_i=A_i\hat{e}_i^{-1}+(1-A_i)(1-\hat{e}_i)^{-1}$. Then, the IPTW estimator of the $CMR$ is 
\begin{equation}
\label{eq:CRRMSM}
\left. {\widehat{CMR}_{IPTW}=\frac{\sum_{i=1}^n \hat{W}_i Y_i A_i}{\sum_{i=1}^n \hat{W}_i A_i}} \middle/ {\frac{\sum_{i=1}^n \hat{W}_i Y_i (1-A_i)}{\sum_{i=1}^n \hat{W}_i (1-A_i)}} \right.\end{equation} 
The estimator (\ref{eq:CRRMSM}) is equal to $\exp(\hat{\beta}_1)$, where $\hat{\beta}_1$ is the weighted least squares estimator of the exposure coefficient when regressing $Y$ on $A$ with weights $\hat{W}$ and a log link. 

If the assumed $A \mid L$ weight model is correctly specified, then (\ref{eq:CRRMSM}) is consistent and asymptotically normal with asymptotic variance $\Sigma_{IPTW}^*$, which can be consistently estimated with the empirical sandwich variance estimator as discussed in Section \ref{sec:methodsvar}. Alternatively, if the weights are known functions of $A$ and $L$, then (\ref{eq:CRRMSM}) is consistent and asymptotically normal with asymptotic variance $\Sigma_{IPTW}$ where $\Sigma_{IPTW}^* \le \Sigma_{IPTW}$. This is analogous to the classic result about the IPTW estimator of the average treatment effect \citep{Lunceford2004}. Note standard statistical software can be used to estimate $\Sigma_{IPTW}$ by the empirical sandwich variance estimator from weighted least squares regression. In practice, the weights are rarely if ever known in the observational setting. The derivations of the asymptotic variance of (\ref{eq:CRRMSM}), both when treating the weights as fixed and when treating the weights as estimated, are included in Section A1 of the Appendix.

\subsubsection{Parametric g-formula}
\label{sec:pargmethods} \citet{robins1986new} introduced the parametric g-formula as a type of standardization that allows for the estimation of causal effects by directly modeling the outcome as a function of the exposure and covariates $L$ and then integrating over the distribution of $L$. The parametric g-formula estimator of the $CMR$ is \begin{equation}
\label{eq:Parametricgform}
\widehat{CMR}_{PG}= \frac{\sum_{i=1}^n \hat{E}(Y_i \mid L_i,A_i=1)}{\sum_{i=1}^n \hat{E}(Y_i \mid L_i,A_i=0)}
\end{equation}  where $\hat{E}(Y_i \mid L_i,A_i=a)$ is computed for $a \in \{0,1 \}$ from the MLEs of the parameters for an assumed model for $Y \mid L,A$. If the assumed parametric model is correctly specified, then (\ref{eq:Parametricgform}) is consistent and asymptotically normal (see proof in Section A2 of the Appendix).

Count outcomes are commonly modeled using the Poisson, negative binomial (NB), zero-inflated Poisson (ZIP), and zero-inflated negative binomial (ZINB) distributions. These distributions can be used to model $Y \mid L,A$ which allows for computation of $\hat{E}(Y_i \mid L_i,A_i=a)$ for $a \in \{0,1 \}$ in (\ref{eq:Parametricgform}). For the Poisson and NB distributions, a generalized linear model (GLM) is fit of the form: $\log(\mu_i)=X_i \gamma$ for $i=1,...,n$, where $\mu_i=E(Y_i \mid A_i, L_i)$ and $X_i=g(L_i,A_i)$ is a vector of predictors for participant $i$ for some user-specified function $g$ of $L_i$ and $A_i$, and $\gamma$ is a column vector of regression coefficients. The MLE $\hat{\gamma}$ is obtained for the GLM and $\hat{E}(Y_i \mid L_i,A_i=a)=\exp\{g(L_i,a) \hat{\gamma}\}$ is calculated for each participant for $a \in \{0,1 \}$.

The ZIP and ZINB distributions account for excess zeros in the count outcome without and with overdispersion, respectively, by assuming that only a portion of the population is susceptible to having a non-zero count while the remaining are not \citep{mullahy1986specification}. When the outcome follows a ZIP or ZINB distribution, models for the probability of individual $i$ not being susceptible ($\nu_i$) and the expected count for individual $i$ within the susceptible population ($\eta_i$) are simultaneously fit: $\mbox{logit}(\nu_i)=X_{i1}\gamma_1$ and $\log(\eta_i)=X_{i2} \gamma_2$, where $X_{i1}=g_1(L_i,A_i)$ and $X_{i2}=g_2(L_i,A_i)$ for user-specified functions $g_1$ and $g_2$, and $\gamma_1$ and $\gamma_2$ are corresponding column vectors of regression coefficients. MLEs $\hat{\gamma_1}$ and $\hat{\gamma_2}$ are obtained for the model and $\hat{E}(Y_i \mid L_i,A_i=a)= [1- \mbox{logit}^{-1}\{g_1(L_i,a)\hat{\gamma_1}\}] \exp\{g_2(L_i,a) \hat{\gamma_2}\}$ is calculated for each participant for $a \in \{0,1\}$. For the ZIP and ZINB models above, regression coefficients have latent interpretations; alternatively, marginalized ZIP and ZINB models may be assumed where parameters have marginal interpretations \citep{long2014marginalized,preisser2016marginalized}. \cite{albert2014estimating} propose and apply (\ref{eq:Parametricgform}) to estimate the $CMR$ assuming a ZINB or zero-inflated beta binomial parametric model for the outcome. 

\subsubsection{Doubly Robust Estimation}
\label{sec:DRmethods}
Next consider doubly robust estimators which incorporate both IPTW and parametric g-formula estimators and are consistent when either the weight or outcome model, but not necessarily both, are correctly specified \citep{bang2005doubly,Hernan2010,FunkDR}. Specifically, the following is a doubly robust estimator for the $CMR$: \begin{equation} \label{eq:DRest}
    \widehat{CMR}_{DR}= \hat{\lambda}^1_{DR} \big/ \hat{\lambda}^0_{DR}
\end{equation} where $\hat{\lambda}^1_{DR}=n^{-1}\sum_{i=1}^n  \hat{e}_i^{-1}\{A_iY_i-(A_i-\hat{e}_i)m_1(L_i,\hat{\gamma})\}$ and $\hat{\lambda}^0_{DR}=n^{-1}\sum_{i=1}^n  (1-\hat{e}_i)^{-1}\{(1-A_i)Y_i+(A_i-\hat{e}_i)m_0(L_i,\hat{\gamma})\}$, $\hat{e}_i$ is the estimated propensity score for participant $i$ from the weight model as described in Section \ref{sec:MSMmethods}, and $m_a(L_i,\hat{\gamma})=\hat{E}(Y_i \mid L_i,A_i=a)$ is the predicted potential outcome for participant $i$ for $a \in \{0,1\}$ from the outcome model, based on the Poisson, NB, ZIP, or ZINB distribution. The causal mean estimators $\hat{\lambda}^a_{DR}$ for $a \in \{0,1\}$ are of the form considered in \citet{Lunceford2004} that were originally proposed by \citet{robins1994estimation}. When either the weight or outcome model is correctly specified, (\ref{eq:DRest}) is consistent and asymptotically normal (see proof in Section A3 of the Appendix).

\subsection{Estimators: Data Heaping}
\label{sec:heapingmethods}
In many settings, the true outcome of interest $Y$ is measured with error. Count outcomes are particularly susceptible to a type of measurement error known as data heaping, which can occur when some participants round or approximate their reported count outcomes rather than reporting exact counts. For example, self-reported cigarette counts are frequently rounded to the nearest multiple of 10 or 20 \citep{klesges1995self,wang2008modeling}. When data heaping is present, statistical methods which ignore heaping will in general not lead to valid inference \citep{wang2008modeling}. 

Consider the following heaping model. First, let $Y_{hi}^a$ denote the heaped potential outcome for participant $i$, i.e., the outcome they would report if, possibly counter to fact, they received treatment $a$. Define the observed (heaped) count as $Y_{hi}=A_iY_{hi}^1+(1-A_i)Y_{hi}^0$. Because $Y_i^a \ne Y_{hi}^a$ for some $i$, $E(Y_{hi}^a) \ne \lambda^a$ in general. Suppose that some individuals report heaped outcomes and other individuals report their outcome exactly. In particular, suppose $Y_{hi}^a=\Delta_i Y_{i}^a+(1-\Delta_i)h_{\eta}(Y_{i}^a)$ for $a \in \{0,1\}$, where $\Delta_i$ is 1 if participant $i$ reported the outcome exactly and 0 otherwise, and $h_{\eta}(Y_{i}^a)$ is a function which rounds $Y_{i}^a$ to the nearest multiple of the known constant $\eta$. Under this model, two participants may report the same value of the outcome for different reasons. For example, consider two smokers who are asked to report the number of cigarettes they smoked the previous day. One woman recalls the exact number, i.e., $\Delta_i=1$, and reports $Y_{hi}=20$. Another woman does not recall the exact number, i.e., $\Delta_i=0$, but estimates approximately one pack of cigarettes and reports $Y_{hi}=20$. Note in the latter setting, it is possible that $Y_i=Y_{hi}=20$. Because $\Delta_i$ is unobserved, one cannot distinguish between these two cases from the observed data.

Below consistent estimators of $CMR$ are presented which allow for outcome heaping. The estimators all take the general form of ``correcting'' a naive estimator that ignores data heaping. In Section \ref{sec:HCAR}, IPTW, parametric g-formula, and doubly robust estimators are given assuming ``heaping completely at random (HCAR),'' i.e., that $\Delta \perp Y$. In Section \ref{sec:infheap}, an an informative heaping estimator is proposed that allows the probability of reporting an exact count to vary depending on the (unobserved) values of the true outcome $Y$.

\subsubsection{Heaping Completely at Random} \label{sec:HCAR}
When data are HCAR, the probability of reporting an exact count is independent of the outcome. Under this assumption it is possible to construct consistent estimators by correcting or adjusting naive estimators that ignore data heaping. Specifically, Section A4 of the Appendix shows that if $\Delta \perp Y$, then $E(Y^a)=\pi^{-1} [ E(Y_h^a)-(1-\pi)E\{h_{\eta}(Y^a)\}]$ for $a \in \{0,1\}$, where $\pi=Pr(\Delta=1)$. Note because $h_{\eta}(Y^a_{i})=h_{\eta}(Y^a_{hi})$ under the assumed heaping model, $E\{h_{\eta}(Y^a)\}=E\{h_{\eta}(Y_h^a)\}$. Therefore the parameters $E(Y_h^a)$ and $E\{h_{\eta}(Y^a)\}$ are identifiable under the assumptions in Section \ref{subsec:prelim}. This result motivates the class of HCAR plug-in estimators for the $CMR$

\begin{equation}{ \label{eqn:HCARplugin}
\widehat{CMR}_{HCAR}=\hat{E}_h(Y^1) \big/ \hat{E}_h(Y^0)}
\end{equation} where $\hat{E}_h(Y^a)=\hat{\pi}^{-1} [ \hat{E}(Y_h^a)-(1-\hat{\pi})\hat{E}\{h_{\eta}(Y^a)\}]$. Note the corrected estimator $\hat{E}_h(Y^a)$ can be rewritten as the naive estimator $\hat{E}(Y_h^a)$ plus a correction term, i.e., $\hat{E}_h(Y^a)=\hat{E}(Y_h^a) + c$ where $c=(1-\hat{\pi})\hat{\pi}^{-1}[\hat{E}(Y_h^a)- \hat{E}\{h_{\eta}(Y^a)\}]$.  When $\hat{\pi}$ is close to one, i.e., according to the fitted model heaping is unlikely, the corrected estimator will be approximately equal to the naive estimator. 

The IPTW, g-formula, and doubly robust estimators based on (\ref{eqn:HCARplugin}) are outlined below. Each of the estimators can be computed by fitting finite dimensional parametric models for $Y \mid Z$, where $Z$ is defined below for each method. Let $f(y;\delta)$ denote the probability mass function (PMF) for the conditional distribution of $Y \mid Z$ with parameter vector $\delta$. Note $f(y;\delta)$ depends on $Z$ but this is left implicit for notational simplicity. Define $l(Y_{hi})=\pi f(Y_{hi};\delta)+(1-\pi) \sum_{y:h_{\eta}(y)=Y_{hi}} f(y; \delta)$. Then, conditional on $Z$, the log-likelihood is proportional to $\mathscr{L} = \sum_{i=1}^n \log l(Y_{hi})$.

\textit{Inverse Probability of Treatment Weighting.}
For the IPTW estimator $\widehat{CMR}_{HCAR,IPTW}$, the propensity score $e_i$ is estimated for each participant and estimated weights $\hat{W}_i$ are constructed, as outlined in Section \ref{sec:MSMmethods}. Then, weights are applied to estimate $E(Y_h^a)$ by
\begin{equation} \label{est:IPTWheap}
    \hat{E}_{IPTW}(Y_h^a)=\frac{\sum_{i=1}^n \hat{W}_i Y_{hi} I(A_i=a)}{\sum_{i=1}^n \hat{W}_i I(A_i=a)}
\end{equation} 
The parameter $E\{h_{\eta}(Y^a)\}$ can be estimated analogously, replacing $Y_{hi}$ with $h_{\eta}(Y_{hi})$ in (\ref{est:IPTWheap}). Let $\hat{\pi}_{marg}$ be the estimator of $\pi$ that maximizes $\mathscr{L}$ under the assumed PMF of $Y$, where $Z$ is the empty set. Then, (\ref{eqn:HCARplugin}) is constructed by plugging in $\hat{\pi}_{marg}$, $\hat{E}_{IPTW}(Y_h^a)$, and $\hat{E}_{IPTW}\{h_{\eta}(Y^a)\}$ for $a \in \{0,1\}$.

\textit{Parametric g-formula.} The parametric g-formula estimator of $CMR$ can be modified to accommodate data heaping under HCAR by replacing the log-likelihood function for $Y \mid A, L$ with $\mathscr{L}$ where $f(y;\delta)$ is the PMF for the assumed $Y \mid A,L$ parametric model, i.e., $Z=\{A, L\}$. The MLEs for the parameters in the heaping model are used to calculate $\hat{Y}^a_{i}=\hat{E}(Y_{i} \mid L_i,A_i=a)$ for each participant, and the PG estimator is constructed as 
\begin{equation}
\label{eq:PGheaping}
\widehat{CMR}_{HCAR,PG}= \frac{\sum_{i=1}^n \hat{Y}^1_{i}}{\sum_{i=1}^n \hat{Y}^0_{i}}
\end{equation}

\textit{Doubly Robust Estimation.} The doubly robust estimator $\widehat{CMR}_{HCAR,DR}$ is constructed from the estimated propensity scores $\hat{e}_i$ used in the IPTW estimator and estimated heaped potential outcomes from the parametric g-formula heaping model. Specifically, $\hat{Y}^a_{hi}=\hat{\pi}_{PG}\hat{Y}^a_{i}+(1-\hat{\pi}_{PG})h_{\eta}(\hat{Y}^a_{i})$, where $\hat{\pi}_{PG}$ is the parametric g-formula estimator of $\pi$. Define 

\begin{equation*}{
\hat{E}_{DR}(Y_h^1)=n^{-1}\sum_{i=1}^n  \hat{e}_i^{-1}\{A_iY_{hi}-(A_i-\hat{e}_i)\hat{Y}^1_{hi}\} }
\end{equation*} and
\begin{equation*}{
\hat{E}_{DR}(Y_h^0)=n^{-1}\sum_{i=1}^n  (1-\hat{e}_i)^{-1}\{(1-A_i)Y_{hi}+(A_i-\hat{e}_i)\hat{Y}^0_{hi}\}}
\end{equation*} 
Then, $\hat{E}_{DR}\{h_{\eta}(Y^a)\}$ is computed analogously for $a \in \{0,1\}$, replacing  $Y_{hi}$ with $h_{\eta}(Y_{hi})$ and $\hat{Y}^a_{hi}$ with $h_{\eta}(\hat{Y}^a_{i})$. While $\hat{\pi}_{PG}$ is computed from the parametric g-formula model and used to estimate $Y_{hi}$ and $h_{\eta}(Y_{hi})$, to ensure double robustness under misspecification of the outcome model, the g-formula estimator of $\pi$ is not plugged into (\ref{eqn:HCARplugin}). Instead, a separate marginal heaping model is specified for estimation of $\pi$, as described above for the IPTW estimator. Then, $\hat{\pi}_{marg}$ from the marginal heaping model, $\hat{E}_{DR}(Y_h^a)$, and $\hat{E}_{DR}\{h_{\eta}(Y^a)\}$ for $a \in \{0,1\}$ are plugged into (\ref{eqn:HCARplugin}) for computation of $\widehat{CMR}_{HCAR,DR}$. Assuming the marginal heaping model used to compute $\hat{\pi}_{marg}$ is correctly specified, then $\widehat{CMR}_{HCAR,DR}$ is a consistent estimator of $CMR$ if either the weight model for computing $\hat{e}_i$ or the heaping outcome model for computing $\hat{Y}^1_{i}$ and $\hat{Y}^0_{i}$ are correctly specified, but not necessarily both. 

\subsubsection{Informative Heaping} \label{sec:infheap} 
The estimators described in Section \ref{sec:HCAR} are consistent estimators of $CMR$ if the HCAR assumption holds. However, often heaping is informative, with the probability of reporting an exact count dependent upon the (unobserved) true count. For example, participants in the WIHS with larger cigarette counts tended to report counts at multiples of 10 more often than participants with smaller cigarette counts. Similar heaping behavior has been observed in other studies of smokers \citep{klesges1995self}.

Here an extension of the parametric g-formula estimator is given that can accommodate informative heaping. Recall under the assumed heaping model that $Y_{hi}^a=\Delta_i Y_{i}^a+(1-\Delta_i)h_{\eta}(Y_{i}^a)$ for $a \in \{0,1\}$. Corresponding to the function $h_{\eta}$, 
define heaping intervals $[0, c_1), [c_1, c_2), ...,  [c_{J-1},c_J)$ such that 
$h_{\eta}(y)$ is constant for all $y \in [c_{j-1}, c_j)$. In other words, the heaping intervals are the level sets of $h_{\eta}$.  The number of heaping intervals $J$ is chosen such that the largest observed $Y$ is contained in $[c_{J-1},c_J)$. For example, suppose that reported cigarette counts are rounded to the nearest 10 under the assumed heaping model and that the largest reported cigarette count is 23. Then the heaping intervals would be $[c_1,c_2)=[0,5)$, $[c_2,c_3)=[5,15)$, and $[c_3,c_4)=[15,25)$.

Assume that data are HCAR within each heaping interval, i.e., that $\Delta \perp Y \mid c_{j-1} \leq Y < c_j$ for $j=1,..,J$. That is, within a heaping interval $j$, the probability of reporting an exact count $\pi_j=Pr(\Delta=1 \mid c_{j-1} \leq Y < c_j)$ is assumed to be the same, but this probability is allowed to differ across heaping intervals such that $\pi_j \ne \pi_k$ in general for $j \ne k$. Such an assumption may be plausible in many applications. For example, it may be reasonable to assume that WIHS participants with true cigarette counts of $8$ and $12$ are equally likely to report a count of $10$, but that the probability of reporting the exact count may be different (e.g., lower) for a participant with a true count of $38$.

When data are HCAR within the heaping intervals, the $CMR$ can be consistently estimated by the parametric g-formula. In particular, the log-likelihood for $Y | A, L$ is now proportional to $\mathscr{L} = \sum_{i=1}^n \sum_{j=1}^J I(c_{j-1} \leq Y_{hi} < c_j) \log l_j(Y_{hi})$, where 
$l_j(Y_{hi})=\pi_j f(Y_{hi};\delta)+(1-\pi_j) \sum_{y:h_{\eta}(y)=Y_{hi}} f(y; \delta)$. Note the parameters $\delta$ of the outcome models are assumed to be shared across heaping intervals such that only the $\pi_j$ parameters are estimated separately across intervals. As in the HCAR setting, the MLEs for the parameters in the heaping model are used to calculate the predicted outcome $\hat{Y}^a_{IH,i}=\hat{E}(Y_{i} \mid L_i,A_i=a)$ for each participant had possibly counter to fact they received treatment $a$. The g-formula estimator is then given by 
\begin{equation}
\label{eq:PGIH}
\widehat{CMR}_{IH,PG}= \frac{\sum_{i=1}^n \hat{Y}^1_{IH,i}}{\sum_{i=1}^n \hat{Y}^0_{IH,i}}
\end{equation}

\subsection{Variance Estimation and Confidence Intervals}
\label{sec:methodsvar}
Each of the proposed $CMR$ estimators from Sections \ref{sec:methodsnoheap} - \ref{sec:heapingmethods} can be expressed as solutions to a vector of unbiased estimating equations (see Appendix sections A1-A3), and therefore are consistent and asymptotically normal under certain regularity conditions \citep{Stefanski2002}. In addition, the asymptotic variance of these estimators can be consistently estimated using the empirical sandwich variance estimator, which can  in turn be used to construct Wald type confidence intervals (CIs).

\section{Simulation Study}
\label{sec:counts-sim}
Simulation studies were conducted to examine and compare the empirical properties of the IPTW, parametric g-formula, and doubly robust estimators of $CMR$ proposed in Section \ref{sec:counts-methods} for a binary exposure $A$ and a count outcome $Y$ in the presence of covariates $L$. Simulations were conducted both without data heaping, where the true outcome was observed (Section \ref{sec:counts-sim-noheap}), and with data heaping, where the observed count was rounded to the nearest ten for some participants (Section \ref{sec:counts-sim-heap}). Within the data heaping setting, both HCAR and informative heaping were considered. For each simulation scenario, 5000 simulated samples were generated and analyzed.

\subsection{Without Data Heaping}
\label{sec:counts-sim-noheap}
The first set of simulations were designed based on the motivating example in Section \ref{sec:counts-WIHS-Partners}, which aimed to estimate the effect of incarceration on the number of sexual partners in a six-month period, controlling for covariates such as age, drug use, and sex exchange practices. A sample of $n=800$ participants was simulated. Simulations were also conducted for $n=2000$, with the results presented in the Appendix. Three covariates $L_1$, $L_2$, and $L_3$ were generated. Representing a participant's baseline age, $L_1$ was simulated from Uniform(20, 40). The covariate $L_2$ represented baseline drug use status and was Bernoulli with mean $\mbox{logit}^{-1}(0.08-L_1/100)$, and $L_3$ represented the baseline sex exchange variable and was Bernoulli with mean $\mbox{logit}^{-1}(-2.9-L_1/100+1.2L_2)$. The exposure $A$ represented the binary incarceration status at the visit following baseline and was Bernoulli with mean $\mbox{logit}^{-1}(-0.5-L_1/100+0.5L_2+0.5L_3)$. 

The outcome of interest $Y$ represented the number of total male sexual partners in the six-month period following measurement of the exposure, with $Y^a \mid L$ generated under the four assumed parametric distributions: Poisson, NB, ZIP, and ZINB. The parameters of the four distributions from Section \ref{sec:pargmethods} equaled $\mu^a=\eta^a=\exp(-1-0.005L_1+0.7L_2+3.5L_3+0.5a)$ and $\nu^a=\mbox{logit}^{-1}(-2.5+L_1/100-0.3L_2-2L_3)$, where superscripts denote the values of parameters under exposure $a \in \{0, 1\}$. The dispersion parameter $\theta=0.5$ was defined such that $Var(Y_i \mid A_i, L_i)=\mu_i+\mu_i^2\theta$ for the NB distribution. For each scenario, $\mbox{log}(CMR)=0.5$.

The estimated causal mean ratios $\widehat{CMR}_{IPTW}$, $\widehat{CMR}_{PG}$, and $\widehat{CMR}_{DR}$ and their estimated variances were calculated for each scenario both under correct model specification and when the weight and/or outcome model were incorrectly specified by excluding $L_2$. Standard errors for $\widehat{CMR}_{IPTW}$ were estimated both conservatively treating the weights as fixed or known, and appropriately treating the weights as estimated. Standard error estimates were computed using the geex package in R \citep{saul}. Corresponding 95\% Wald confidence intervals (CIs) were computed throughout. The maximum estimated Monte Carlo standard error \citep{morris2019using} for the mean percent bias across estimators and scenarios was 0.26\%, and the Monte Carlo standard error for coverage at the nominal level was 0.31\%. 

The results of the simulation for $n=800$ are presented in Figure \ref{fig:SimsNoHeap}, with more detailed results in Appendix Tables A1 and A2. These results demonstrate minimal empirical bias regardless of the method or underlying distribution of the data when models were correctly specified. Empirical bias was even smaller when the sample size was increased to $n=2000$ (see Appendix Table A3). For the IPTW estimator, empirical coverage was close to the nominal 95\% level when the weight model was correctly specified and weights were treated as estimated, but was at or near $100\%$ when weights were treated as known. This aligns with the inflated median estimated standard error (MSE) relative to the empirical standard error (ESE) when the weights are treated as known, resulting in the standard error ratio (SER) being above one. The IPTW estimator with weights treated as estimated, parametric g-formula, and doubly robust estimators all had SERs close to one, demonstrating the consistency of the empirical sandwich variance estimator. The parametric g-formula and doubly robust estimators yielded more precise estimates than IPTW, with the parametric g-formula having the smallest MSEs (Appendix Table A1). 

As anticipated, the doubly robust estimators yielded minimal bias when either the weight or outcome model was correctly specified, while the IPTW and parametric g-formula estimators were biased under weight and outcome model misspecification, respectively (see Figure \ref{fig:SimsNoHeap}, Appendix Table A2, and Appendix Table A4). The doubly robust estimators were biased when both models were misspecified. For the doubly robust estimators, MSEs were smaller when the outcome model was correctly specified than when it was misspecified; the MSEs were similar when the weight model was misspecified compared to correctly specified (Appendix Table A2 and Appendix Table A4). These findings are consistent with the empirical results in \citet{FunkDR}. 

ZIP and ZINB models failed to converge for between 0.3\% and 3.5\% of simulations when $n=800$, and between 0\% and 1.2\% of simulations when $n=2000$ (Appendix Tables A3 - A4). This amount of non-convergence is in line with empirical findings from other studies using mixture models \citep{preisser2016logistic, benecha2017marginalized}.

\begin{figure}
\begin{center}
  \includegraphics[clip, trim=0.5cm 0cm 0cm 0cm, width=\linewidth]{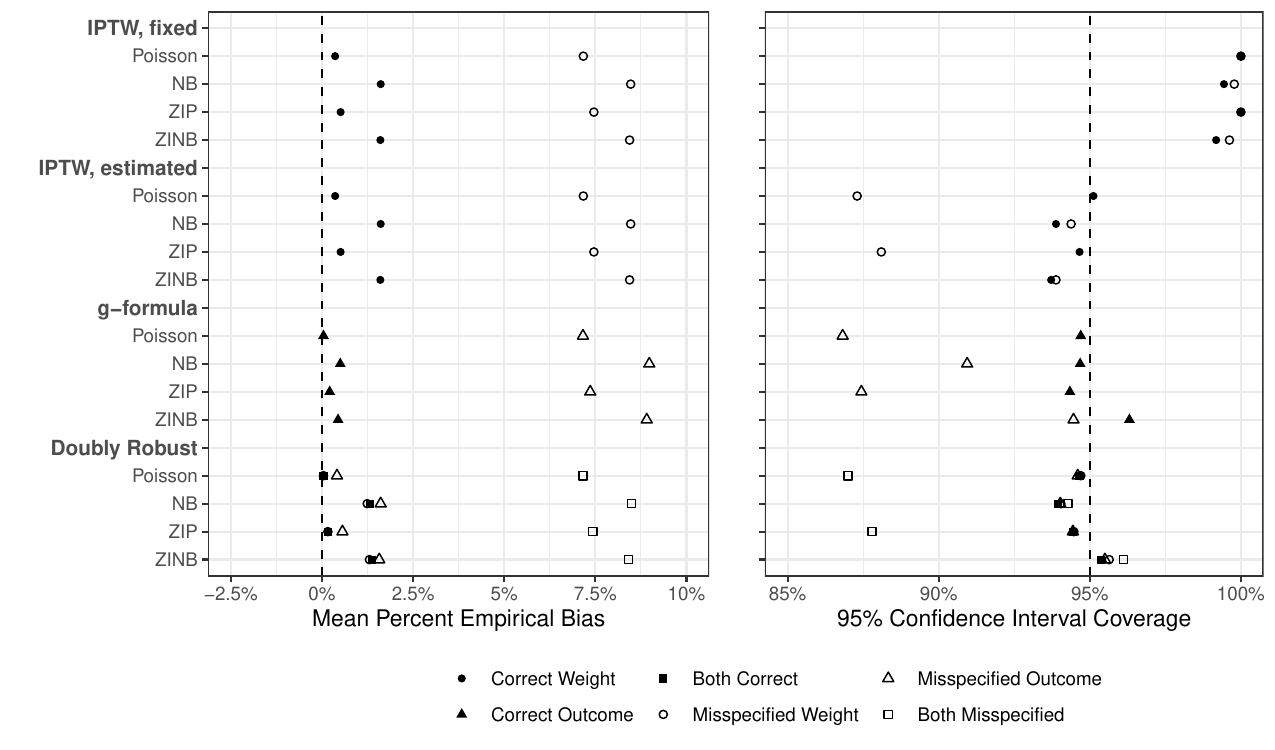}
  \caption{Results of the simulation study by method and distribution across $5000$ samples with correct and incorrect model specification, $n=800$, without data heaping. Percent empirical bias and 95\% confidence interval coverage calculated for the causal mean ratio.}
  \label{fig:SimsNoHeap}
\end{center}
\end{figure}

\subsection{With Data Heaping}
\label{sec:counts-sim-heap}

To demonstrate the empirical properties of the estimators that account for data heaping, data were simulated where the outcome was heaped, such that the estimators from Section \ref{sec:methodsnoheap} were expected to be biased. This simulation study was designed based on the motivating example in Section \ref{sec:counts-WIHS-Cigs}, which aimed to estimate the effect of incarceration on the number of cigarettes smoked per day among smokers in a six-month period, controlling for covariates such as income. As with the simulations without data heaping, a sample of $n=800$ participants was simulated, with additional simulations conducted for $n=2000$ presented in the Appendix. The covariate $L_4$ was simulated such that $\exp(L_4) \sim Gamma(a=5, s=2)$ where the density function for the Gamma distribution was $f(x)= \{{s}^{a}\Gamma(a)\}^{-1} {x}^{a-1} \exp(-x/s)$, and $\Gamma(z)=\int_0^\infty t^{z-1} \exp(-t) dt$. The exposure $A$ was simulated from a Bernoulli distribution with mean $1-\mbox{logit}^{-1}(-0.8+0.65L_4)$. The potential outcomes for the number of cigarettes smoked under exposure and no exposure were simulated as $Y^a \mid L_4 \sim Poisson(\mu^a)$ for $a \in \{0,1\}$, where  $\mu^a=\exp(-0.9+L_4+0.25a)$. Thus, $\mbox{log}(CMR)=0.25$. Under this data generating mechanism, the conditional distribution of $Y^a \mid L$ was Poisson and the marginal distribution of $Y^a$ was NB. 

Data heaping were induced with $\eta=10$ and $\Delta_i$ simulated from a Bernoulli distribution. In Scenario 1 (HCAR), $E(\Delta)=\pi=0.4$ for all observations. In Scenario 2 (informative heaping), $E(\Delta \mid Y)=\pi_j$, with $\pi_1=0.5$ for $Y<5$ and $\pi_2=0.2$ for $Y \geq 5$, such that the probability of reporting the exact count was more likely for smaller counts compared to larger counts. Under this data generating mechanism, $\pi_2=\pi_3=...=\pi_J$, so only two heaping intervals were assumed in estimation. True and heaped counts for a single iteration of the simulation are presented in Appendix Figure A1. 

The estimated $CMRs$ based on the IPTW, parametric g-formula, and doubly robust estimators and their estimated variances were calculated for each scenario ignoring data heaping, using the estimators described in Section \ref{sec:methodsnoheap} (referred to as naïve in the results below), and accounting for data heaping using the methods described in Section \ref{sec:heapingmethods}. To assess the sensitivity of the heaping estimators to the HCAR assumption, both the HCAR estimators and informative heaping parametric g-formula estimator were applied in Scenarios 1 and 2. As with the simulations presented in Section \ref{sec:counts-sim-noheap}, the $CMR$ was estimated under correct model specification and when the weight and/or outcome model were incorrectly specified. For incorrectly specified models, the covariate $L_5$ was included in the model(s) instead of $L_4$, where $L_5=\mbox{logit}^{-1}(-3+L_4+2\epsilon_3)$ and $\epsilon_3$ is simulated from a standard normal distribution. The maximum estimated Monte Carlo standard error for the mean percent bias across estimators and scenarios was 0.10\%, and the Monte Carlo standard error for coverage at the nominal level was 0.31\%.

The results of the data heaping simulations are presented in Figure \ref{fig:SimsHeap}, with more detailed results in Appendix Table A5 and Appendix Table A7. When data heaping was accounted for under the appropriate heaping assumption (HCAR or informative heaping), the results were similar to those presented in Section \ref{sec:counts-sim-noheap}. That is, the heaping estimators had low empirical bias and close to nominal CI coverage under correct model specification. The doubly robust HCAR estimator demonstrated low bias under incorrect specification of one (but not both) of the weight or outcome models in Scenario 1, but was biased in Scenario 2 when the HCAR assumption was violated. When data heaping was ignored and the naïve estimators from Section \ref{sec:methodsnoheap} were applied to heaped data, the estimates exhibited considerable bias and 95\% CI coverage was below the nominal level (Appendix Table A5 and Appendix Table A7). Similarly, estimators that assumed HCAR were generally biased in Scenario 2, when heaping was informative. However, the parametric g-formula HCAR estimator was robust to violation of the HCAR assumption under the data generating mechanism considered, with similar performance to the parametric g-formula informative heaping estimator.

\begin{figure}
\begin{center}
  \includegraphics[clip, trim=0cm 0cm 0cm 0cm, width=\linewidth]{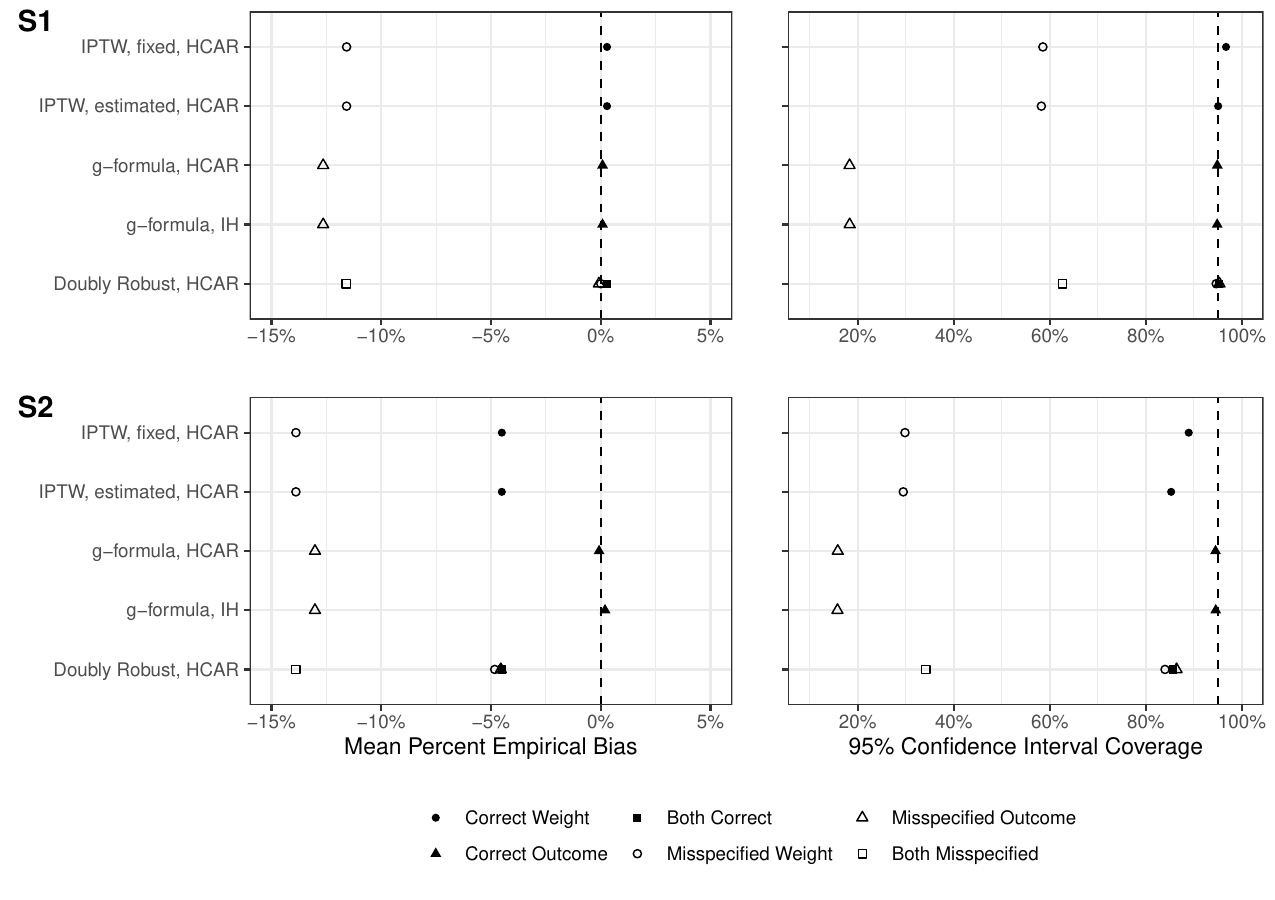}
  \caption{Results of the data heaping simulation study for Scenario 1 (S1), Heaping Completely at Random (HCAR), top, and Scenario 2 (S2), Informative Heaping (IH), bottom, by method across $5000$ samples with correct and incorrect model specification, $n=800$. Percent empirical bias and empirical 95\% CI coverage calculated for the $CMR$. Results exclude one simulation per scenario where models did not converge.}
  \label{fig:SimsHeap}
\end{center}
\end{figure}

\section{Women's Interagency HIV Study Analysis}
\label{sec:counts-WIHS}
The methods from Section \ref{sec:counts-methods} were applied to data from the WIHS to estimate the effect of incarceration in the past six months on two outcomes in the subsequent six months: the number of male sexual partners (Section \ref{sec:counts-WIHS-Partners}) and the number of cigarettes smoked per day among smokers (Section \ref{sec:counts-WIHS-Cigs}).   

\subsection{Number of Partners}

\label{sec:counts-WIHS-Partners}
The WIHS sample used to estimate the effect of incarceration in the past six months on the total number of male sexual partners (subsequently referred to as partners) during the following six-month period was created by restricting the longitudinal WIHS data set of 4,982 women to women who attended at least one visit between 2007-2017, as 2007 is when incarceration questions were added to the WIHS questionnaire. The data set was further restricted to include only women without missing covariates following implementation of last value carried forward and next value carried back imputation, excluding the history of incarceration covariate which was only asked at a single timepoint and thus could not be imputed using this method. For each woman who reported being incarcerated between 2007-2017, her first incarcerated visit following a non-incarcerated visit was selected as her baseline visit. This allowed for an appropriate run-in period in which to measure covariates at the visit preceding baseline. The outcome was measured at the visit following baseline. This resulted in $n=294$ incarcerated women after excluding the 28 women missing outcome data at the visit following baseline. A sample of one visit from each of $n=588$ women who did not report being incarcerated between 2007-2017 was randomly selected. The sample of unincarcerated women was restricted to women with non-missing outcome data at the visit following baseline and was stratified by visit number to ensure the same distribution of baseline visits over calendar time as the incarcerated women. Unincarcerated women selected for an earlier WIHS visit were not eligible to be selected at a later WIHS visit. Missing values for prior incarceration for $n=14$ participants were imputed with the mode (no history of incarceration). The resulting WIHS sample consisted of 882 women, 68\% of whom had HIV.  At the visit prior to baseline, incarcerated women reported a mean of 1.7 partners (Standard Deviation (SD): 4.2) in the previous six-month period, while those not incarcerated reported a mean of 0.9 partners (SD: 3.4) in the previous six-month period.

\begin{figure}
\begin{center}
  \includegraphics[width=\linewidth]{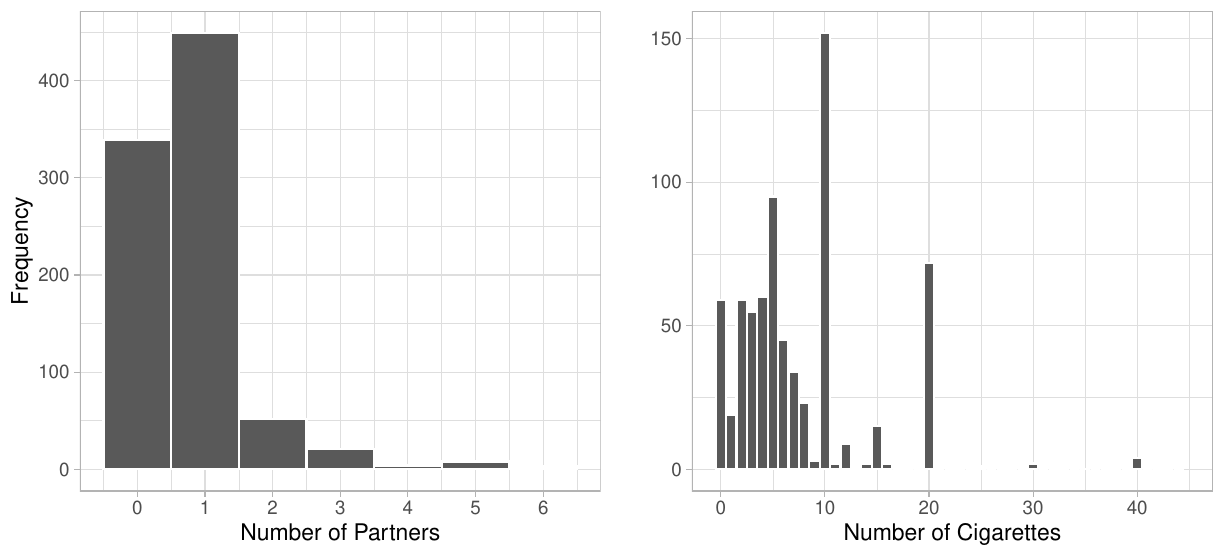}
  \caption{Distribution of male sexual partners (left, $n=882$) and cigarettes smoked per day among smokers (right, $n=716$) during the six months following baseline reported by Women's Interagency HIV Study (WIHS) participants}
  \label{fig:WIHShist}
\end{center}
\end{figure}

It was assumed that potential outcomes were independent of the exposure conditional on the following covariates, measured at the visit prior to baseline: age, educational attainment (high school or more versus less than high school), race (Black, White, or other), six-level collapsed WIHS site (Bronx and Brooklyn, NY; Washington, DC; Los Angeles, CA; San Francisco, CA; Chicago, IL; Southern Sites - Chapel Hill, NC, Atlanta, GA, Miami, FL, Birmingham, AL, and Jackson, MS), HIV status (positive or negative), binary prior incarceration status, unstable housing (living in a rooming/boarding/halfway house versus other housing), sex exchange practices (exchanging sex for drugs, money, or shelter versus not), alcohol use (none, 1-7 drinks/week, or $>$ 7 drinks/week), binary marijuana use, and illicit drug use (use of crack cocaine, cocaine, heroin, methamphetamines, other opioids, or any injection use versus none). There was considerable overlap in the observed distribution of these covariates between incarcerated and unincarcerated participants (Appendix Table A9), though unincarcerated participants were older on average and less likely to have unstable housing, engage in sex exchange practices, or use illicit drugs. Because many participants reported no partners over a six-month period (Figure \ref{fig:WIHShist}), zero-inflated models were considered when estimating the $CMR$ based on parametric g-formula and doubly robust estimation methods. When fitting zero-inflated models to these data, the susceptibility model included the vector of covariates consisting of age, marital status (legally married/common-law married/living with a partner or widowed/divorced/marriage annulled/separated/never married/other), sex exchange practices, HIV status, and sexual orientation (lesbian/gay or heterosexual/straight/bisexual/other), also measured at the visit prior to baseline. These variable classifications were made to predict a woman's potential to have one or more male sexual partners in subsequent study visits. 

The estimators $\widehat{CMR}_{IPTW}$, $\widehat{CMR}_{PG}$, and $\widehat{CMR}_{DR}$ were calculated as described in Section \ref{sec:methodsnoheap} and empirical sandwich variance estimates were computed using the geex package in R. For each estimate, 95\% CIs were constructed as described in Section \ref{sec:methodsvar}. When calculating standard errors for $\widehat{CMR}_{IPTW}$, the weights were treated as estimated in the computation of standard errors. To compare the fit of parametric models to these data, the Akaike information criterion (AIC) was computed for each parametric model. AIC values for the Poisson, NB, ZIP, and ZINB distributions were 2321, 2125, 2281, and 2094, respectively, indicating that the ZINB distribution provided the best fit and was thus used for parametric g-formula and DR estimation. 

Across the three methods, estimated counterfactual means for the number of partners under incarceration and no incarceration ranged from 1.1-1.2 and 0.8-0.9, respectively. The estimated $CMR$ (95\% CIs) for the IPTW, parametric g-formula, and doubly robust methods were 1.27 (0.68, 1.85), 1.34 (0.91, 1.77), and 1.24 (0.70, 1.78), respectively. The expected number of partners if incarcerated is estimated to be about 1.3 times the expected number of partners if not incarcerated, but confidence intervals for all three methods span the null. Precision estimates were similar across methods, with the parametric g-formula having the smallest estimated standard error and the IPTW having the largest estimated standard error. 

Two sensitivity analyses were conducted. Instead of using last value carried forward and next value carried back imputation for missing covariates and excluding women with missing outcome data, multiple imputation was applied to the longitudinal WIHS data, and the analytic sample was derived within each imputed data set. In the first sensitivity analysis, the previously described analysis was repeated in each of the 30 multiply imputed data sets, and results were combined using Rubin's method \citep{rubin2004multiple}. In the second sensitivity analysis, the first sensitivity analysis was repeated with the number of sexual partners at baseline, categorized as $0$, $1$, or $2+$, included as an additional covariate in the weight and/or outcome models for the three estimators. The results of both sensitivity analyses were similar to the primary findings, though less precise. The details and results of the sensitivity analyses are included in Section B1 of the Appendix.

The findings here complement the results in \citet{knittel}, as both studies estimate an increase in the number of sexual partners due to incarceration using WIHS data. \citet{knittel} estimate the effect of incarceration on the categorized number of sexual partners by fitting a generalized logit model with IPTW, controlling for the previously mentioned covariates. In their analysis of the WIHS data, the estimated odds of having 2 sexual partners (versus 1 partner) if incarcerated were 2.41 (95\% CI: 1.20, 4.85) times the odds if not incarcerated. Odds ratios (95\% CIs) for 0 and 3+ sexual partners were 1.20 (0.66, 2.17) and 2.03 (0.97, 4.26), respectively. While useful for demonstrating the direction of the effect, odds ratios are difficult to interpret \citep{norton2018odds}. The estimates of the CMR presented here are easier to interpret and, unlike the logit model analysis, do not require choices regarding the number of outcome categories or which values of the response to collapse.

\subsection{Number of Cigarettes}
\label{sec:counts-WIHS-Cigs}
The $CMR$ for the effect of incarceration in the past six months on the number of cigarettes smoked per day among smokers in the subsequent six-month period was estimated using the methods in Section \ref{sec:heapingmethods}. As shown in Figure \ref{fig:WIHShist}, cigarette counts exhibited overdispersion and data heaping at multiples of ten, and ignoring this measurement error could lead to biased estimates of the $CMR$. The same set of covariates as in Section \ref{sec:counts-WIHS-Partners} was assumed to provide conditional exchangeability for the number of cigarettes outcome as the number of partners outcome, except that the sex exchange variable was removed and household income (\$12,000 per year or less versus $>$ \$12,000) was included. As with the number of partners outcome, the covariates included in the cigarette outcome analysis demonstrated considerable overlap between incarcerated and unincarcerated participants (Appendix Table A10), but again unincarcerated participants were older on average and less likely to have unstable housing or use illicit drugs.

The analytic data set was derived similarly to the data set for the partners outcome, with the previously stated modifications to the covariate set and an added requirement that women were current smokers at the visit preceding baseline. This resulted in $n=179$ incarcerated women with complete outcome data. A stratified (by WIHS visit) sample of $n=537$ women who reported no incarcerations during the study period were selected with the same distributions of baseline visits over calendar time as the incarcerated women, for a total sample size of $n=716$. Missing prior incarcerations were imputed to the mode for $n=6$ participants. Participants could report cigarettes smoked per day in cigarette or pack counts; it was assumed that one pack of cigarettes equated to 20 cigarettes. In the analytic sample, 66\% of women had HIV. At the visit prior to baseline, incarcerated women reported smoking a mean of 9.2 cigarettes per day (SD: 7.5) in the previous six-month period, while those not incarcerated reported smoking a mean of 8.0 cigarettes per day (SD: 6.2) in the previous six-month period.

The reported number of cigarettes smoked per day clearly exhibited data heaping, as shown in Figure \ref{fig:WIHShist}. Both HCAR and informative heaping estimators were applied, where the informative heaping estimator allowed for the possibility that the probability of reporting an exact count varied across the following heaping intervals: $<5$, $5-14$, and $15+$. There were 252, 366, and 98 participants, respectively, in each heaping interval. The NB distribution was used for estimation to account for overdispersion in the reported number of cigarettes. Depending on the estimator employed, the estimated counterfactual means for cigarettes smoked per day under incarceration and no incarceration ranged from 7.0-7.2 and 7.3-7.5, respectively. The estimated $CMR$ (95\% CI) across the four methods were similar, with $\widehat{CMR}_{HCAR,IPTW}=1.04$ $(0.76, 1.32)$, 
$\widehat{CMR}_{HCAR,PG}=1.04$ $(0.88, 1.20)$, $\widehat{CMR}_{HCAR,DR}=1.03$ $(0.74, 1.33)$, and $\widehat{CMR}_{IH,PG}=1.04$ $(0.87, 1.20)$. An estimated $65\%$ of participants reported exact counts based on the three HCAR estimation approaches, where as the informative heaping analysis suggests the probability of reporting an exact count declined across the three heaping intervals, with $\hat{\pi}_1=0.85$, $\hat{\pi}_2=0.64$, and $\hat{\pi}_3=0.21$. Regardless of the heaping estimator used, there is little evidence of an effect of incarceration on the number of cigarettes smoked per day among smokers, as the expected number of cigarettes smoked if incarcerated is estimated to be about the same as the expected number of cigarettes smoked if not incarcerated. In addition, confidence intervals are wide and are consistent with either a decrease or increase in smoking behavior due to incarceration.  

As with the partners outcome, two sensitivity analyses based on multiple imputation were conducted for the number of cigarettes outcome. In the first sensitivity analysis, the previously described analysis was repeated in each of the 30 multiply imputed data sets, and results were combined using Rubin's method. In the second sensitivity analysis, the first sensitivity analysis was repeated with the reported number of cigarettes smoked at baseline, categorized as $<10$ or $10+$, included as an additional covariate in the weight and/or outcome models for the estimators. In both sensitivity analyses, the estimates of the $CMR$ were slightly larger than these primary findings, ranging from $1.09$ to $1.19$, though estimates were imprecise and their corresponding confidence intervals overlapped heavily with the primary findings. The details and results of the sensitivity analyses are included in Section B2 of the Appendix.

\section{Discussion}
\label{sec:counts-discussion}
This paper considers estimators of the causal mean ratio based on marginal structural modeling with IPTWs, the parametric g-formula, and doubly robust estimation. Estimators are developed for outcomes measured without error or subject to data heaping, and each class of estimators accommodates overdispersion or zero-inflation in the outcome. In the absence of measurement error, consistency and asymptotic normality holds for the IPTW and parametric g-formula estimators under correct exposure and outcome model specification, respectively, and for the doubly robust estimators when either the exposure or the outcome model is correctly specified. Consistency of the proposed heaping estimators relies on the additional assumption that the heaping model is correctly specified. 

Simulations demonstrate that all estimators were empirically unbiased under correct model specification and Wald confidence intervals based on the empirical sandwich variance estimator generally had nominal coverage. The parametric g-formula and doubly robust estimators were more precise than the IPTW estimator in the absence of data heaping but required correct specification of the parametric distribution of the outcome. One notable advantage of the IPTW estimator is that it does not require specification of a parametric model for the outcome when data heaping is not present. The IPTW variance estimator was overly conservative when weights were treated as fixed and thus the use of standard software which treats the weights as known is not recommended. Instead, the sandwich variance estimator which accounts for estimation of the weights can be used.  

Applying the methods considered in this manuscript to WIHS data provides insights regarding the effects of incarceration on women with HIV or at risk of acquiring HIV that are important for public health policy and practice. Analysis of WIHS data indicates that incarceration may increase the number of sexual partners in the subsequent six-month period, with the expected number of partners if incarcerated estimated to be 1.3 times the expected number of partners if not incarcerated. Because increases in sexual partners can result in new exposures to and transmissions of sexually transmitted infections, these findings support the need for sexually transmitted infection prevention interventions for incarcerated women. However, these results should be interpreted with caution as confidence intervals for all three methods included the null value. Sensitivity analyses based on multiple imputation were similar to the primary findings.

In contrast with prior research showing increases in smoking during incarceration among a sample of women in prison \citep{cropsey2008smoking}, our findings were less conclusive. We find no evidence of an effect of incarceration on the number of cigarettes smoked per day among smokers in the six-month period following incarceration. Confidence intervals for the causal mean ratio included the null value and were compatible with both increases and decreases in smoking behavior. These findings could potentially be explained by the smoking behavior of WIHS participants prior to baseline and  by the implementation of smoking bans in prisons and jails. Smokers in the WIHS reported high cigarette counts prior to baseline, which limited what increase could feasibly be observed following incarceration. The influence of smoking bans on these findings is difficult to assess, as WIHS participants who experienced incarceration had varying lengths of incarceration and came from a diverse set of states and municipalities with different smoking policies. By 2007, 60\% of prisons in the United States prohibited all smoking among incarcerated persons, but 40\% still allowed smoking in some locations \citep{kauffman2008tobacco}, and jail policies can differ from prison policies. Additional research is needed to examine longer-term effects of incarceration on smoking behavior in this population and to compare effects across subpopulations (e.g., by HIV status, age, region, smoking history, and duration of incarceration). 

The results of both analyses depend on the validity of the causal identification assumptions. The positivity assumption requires that all women in the WIHS have a non-zero risk of incarceration during the study period. While covariate distributions between incarcerated and unincarcerated participants overlapped considerably, there were few unincarcerated participants with unstable housing, engaged in sex exchange practices, or with illicit drug use. For the sexual partners outcome, the conditional exchangeability assumption assumes potential outcomes are independent of incarceration status conditional on age, educational attainment, race, WIHS site, HIV status, prior incarceration status, unstable housing, sex exchange practices, alcohol use, marijuana use, and illicit drug use. These same covariates were included in the cigarette outcome analysis, except that sex exchange practices was replaced with household income. Conditional exchangeability assumptions are violated if there is residual confounding within strata defined by these covariates. Credible causal inference also relies on the treatment or exposure variable being sufficiently well-defined \citep{Hernan2010}, i.e., there are no hidden versions of treatment \citep{neyman1923application,rubin1974estimating}. In the WIHS data analysis, whether or not an individual is incarcerated is clearly well defined. Nonetheless, variations in the incarceration circumstances (e.g., duration, type of facility) were not accounted for in the analysis and therefore the results rely on the additional assumption of treatment-variation irrelevance \citep{vanderweele2009concerning}. Future research could examine the effects of different types of incarceration. 

While motivated by the WIHS data, these methods are applicable for estimating an exposure effect based on observational data in a variety of settings when the outcome of interest is a count, potentially subject to overdispersion or data heaping. The results in this manuscript apply to causal mean ratios, i.e., the ratios of counterfactual mean counts over a fixed period of time. When follow-up time varies across members of the population, an alternative estimand to the causal mean ratio is the causal rate ratio, i.e., the ratio of counterfactual rates under exposure and no exposure. Rates can be estimated using various methods, including similar modeling approaches to those used for counts, but incorporating an offset to account for varied follow-up time. Therefore, it should be straightforward to adapt the methods in this paper to allow for inference about rate ratios in settings where follow-up time varies across individuals. These methods could be further extended to accommodate more complex data heaping structures. Semiparametric or non-parametric methods, e.g., using targeted maximum likelihood estimation or machine learning, could also be considered to relax the parametric modeling assumptions of the estimators considered here. Methods are also needed to estimate causal estimands when the exposure is a count potentially subject to data heaping.

	\bibliographystyle{apalike}
	\bibliography{bibliography}
	
\section*{Acknowledgements}
This research was supported by NIH grants R01 AI085073 and R01 AI157758 and in part through Developmental funding from the University of North Carolina at Chapel Hill Center For AIDS Research (CFAR), an NIH funded program P30 AI050410. The authors thank John Preisser, Shaina Alexandria, Bryan Blette, Kayla Kilpatrick, Jaffer Zaidi, Samuel Rosin, and Paul Zivich for their helpful suggestions. Data in this manuscript were collected by MACS and WIHS, now the MACS/WIHS Combined Cohort Study (MWCCS), which is supported by the National Institutes of Health. Full acknowledgement is provided in the Appendix and at 

\noindent{h}ttps://statepi.jhsph.edu/mwccs/acknowledgements. The authors gratefully acknowledge the contributions of the study participants and dedication of the staff at the MWCCS sites. \vspace*{-8pt}

\section*{Code and Data Availability Statement}
Access to individual-level data from the MACS/WIHS Combined Cohort Study Data (MWCCS) may be obtained upon review and approval of a MWCCS concept sheet. Links and instructions for online concept sheet submission are on the study website (http://mwccs.org/). R code for computing the different estimators along with the corresponding standard error estimators is available at https://github.com/bonnieshook/Causal\textunderscore Inference\textunderscore Count\textunderscore Outcomes.
	
	\clearpage
	
	\begin{appendices}
		\setcounter{equation}{0}
		\renewcommand{\theequation}{A.\arabic{equation}}
\section*{Appendix A: Proofs of Main Results}
\label{s:webA_TechDeriv}
\renewcommand{\thesubsection}{A\arabic{subsection}}
\renewcommand{\theequation}{A.\arabic{equation}}

\subsection{Section 2.2.1 Derivations}
\label{IPTWresults}

\subsubsection{Derivation of the asymptotic variance of (2) when the weights are treated as fixed}
\label{Prop1WtsFixed}
Let $O_i=(Y_i,A_i,L_i)$ and $\Lambda=(\lambda^0,\lambda^1)$. The estimating equations for $\hat{\lambda}^1_{IPTW}$ and $\hat{\lambda}^0_{IPTW}$ are \begin{equation}\label{EEiptw} \sum_{i=1}^n \psi(O_i, \Lambda) = \begin{bmatrix}
           \sum_{i=1}^n \psi_1(O_i,\Lambda)\\
           \sum_{i=1}^n \psi_0(O_i,\Lambda)\\
            \end{bmatrix} = \begin{bmatrix}
            \sum_{i=1}^n e_i^{-1}(Y_i-\lambda^1)I(A_i=1) \\
           \sum_{i=1}^n (1-e_i)^{-1}(Y_i-\lambda^0)I(A_i=0)\\
            \end{bmatrix} = 0 \end{equation} 
Assuming causal consistency and conditional exchangeability:
\begin{align} 
& E\{\psi_1(O_i,\Lambda)\} = E\{e_i^{-1}(Y_i-\lambda^1)I(A_i=1)\}\nonumber \\
&= E_L \left \{e_i^{-1} E(A_i \mid L_i)E(Y^1_i -\lambda^1 \mid L_i)\right \} \nonumber \\
 &=E(Y^1_i)-\lambda^1=0 \nonumber
\end{align}
Similarly, $E\{\psi_0(O_i,\Lambda)\}=0$. Therefore, (\ref{EEiptw}) is an unbiased set of estimating equations. It follows that under certain regularity conditions \citep{Stefanski2002}, as $n \rightarrow \infty$,
\[\sqrt{n} \begin{bmatrix}
          \hat{\lambda}^1_{IPTW} - \lambda^1 \\
          \hat{\lambda}^0_{IPTW} - \lambda^0
            \end{bmatrix} \xrightarrow[]{d} N \left(0, V(\Lambda) \right)
\]where $V(\Lambda)=A(\Lambda)^{-1} B(\Lambda) \{A(\Lambda)^{-1}\}^T$, $A(\Lambda)=E(-\dot{\psi})$, $B(\Lambda)=E[\psi(O_i, \Lambda) \psi(O_i, \Lambda)^T]$, and 
\[
\dot{\psi}(O_i, \Lambda) = \frac{\partial \psi(O_i, \Lambda)}{\partial \Lambda} = \begin{bmatrix}
          -e_i^{-1} I(A_i=1) & 0 \\
          0 & -(1-e_i)^{-1} I(A_i=0)
            \end{bmatrix}\] 
Note $A(\Lambda)=I_{2 \times 2}$ where $I_{2 \times 2}$ is the identity matrix. By causal consistency, iterated expectation, and conditional exchangeability it is straightforward to show that\[B(\Lambda)=E[\psi(O_i, \Lambda) \psi(O_i, \Lambda)^T] = E \begin{bmatrix}
          e_i^{-2}(Y_i-\lambda^1)^2I(A_i=1) & 0 \\
          0 & (1-e_i)^{-2}(Y_i-\lambda^0)^2 I(A_i=0)
            \end{bmatrix}\]
            \[=E\begin{bmatrix}
          e^{-1}(Y^1-\lambda^1)^2 & 0 \\
          0 & (1-e)^{-1}(Y^0-\lambda^0)^2 \end{bmatrix}\] 
The delta method can then be used to obtain the asymptotic variance of $\widehat{CMR}_{IPTW}=\hat{\lambda}^1_{IPTW}/\hat{\lambda}^0_{IPTW}$. Specifically, let $g(\Lambda)=\lambda^1/\lambda^0$ such that $\partial g / \partial \Lambda = (1 / \lambda^0, -\lambda^1 / (\lambda^0)^2 )$. Then,
\[\sqrt{n} \left( \frac{\hat{\lambda}^1_{IPTW}}{\hat{\lambda}^0_{IPTW}} - \frac{\lambda^1}{\lambda^0} \right) \xrightarrow[]{d} N \left(0, \Sigma_{IPTW} \right)
\]
where
\[\Sigma_{IPTW}=\frac{\partial g}{\partial \Lambda} V(\Lambda)\left(\frac{\partial g}{\partial \Lambda}\right)^T=E \left[ e^{-1} \left( \frac{Y^1-\lambda^1}{\lambda^0}  \right)^2 + (1-e)^{-1} \left\{ \frac{\lambda^1 (Y^0-\lambda^0)}{(\lambda^0)^2}  \right\}^2   \right]
\] 

\subsubsection{Derivation of the asymptotic variance of (2) when the weights are treated as estimated}
When the weights are treated as estimated rather than fixed, consider the set of estimating equations 
\[\sum_{i=1}^n \psi(O_i, \Lambda) = \begin{bmatrix}
          \sum_{i=1}^n \psi_{\alpha}(O_i,\Lambda)\\
          \sum_{i=1}^n \psi_1(O_i,\Lambda)\\
           \sum_{i=1}^n \psi_0(O_i,\Lambda)\\
            \end{bmatrix} = \begin{bmatrix}
          \sum_{i=1}^n \psi_\alpha(O_i,\alpha)\\
          \sum_{i=1}^n W_i(\alpha) (Y_i-\lambda^1)I(A_i=1) \\
           \sum_{i=1}^n W_i(\alpha)(Y_i-\lambda^0)I(A_i=0)\\
            \end{bmatrix} = 0 \]   where the parameter vector $\Lambda^T=(\alpha^T, \lambda^1 , \lambda^0)$ includes the $p$ parameters from the logistic regression weight model ($\alpha$) and the two causal means ($\lambda^1 $ and $\lambda^0 $), and $\psi_{\alpha}$ is the vector of score functions from the logistic regression weight model.
     
Let the solutions to the estimating equations be denoted by $\hat{\Lambda}=[\hat{\alpha},  \hat{\lambda}^1_{IPTW}, \hat{\lambda}^0_{IPTW}]^T$, where $\hat{\lambda}^a_{IPTW}=\sum_{i=1}^n W_i(\hat{\alpha}) Y_iI(A_i=a) / \{\sum_{i=1}^n W_i(\hat{\alpha})I(A_i=a)\}$ for $a \in \{0,1\}$. When the weight model is correctly specified, $\hat{\Lambda}$ is the solution to an unbiased set of estimating equations. Thus, $\sqrt{n}(\hat{\Lambda}-\Lambda) \xrightarrow[]{d} N(0,V(\Lambda))$, where $V(\Lambda)=A(\Lambda)^{-1}B(\Lambda)\{A(\Lambda)^{-1}\}^{T}$, $A(\Lambda)=E\{-\dot{\psi}(O_i,\Lambda)\}$, $B(\Lambda)=E\{\psi(O_i,\Lambda)\psi(O_i,\Lambda)^T\}$, and $\dot{\psi}(O_i,\Lambda)=\partial\psi(O_i,\Lambda)/\partial\Lambda$ are $(p+2) \times (p+2)$ matrices. Also note that:
\[\dot{\psi}(O_i,\Lambda)=\begin{bmatrix}
          \partial\psi_{\alpha}/\partial\alpha & \partial\psi_{\alpha}/\partial\lambda^1  & \partial\psi_{\alpha}/\partial\lambda^0 \\
          \partial\psi_{1}/\partial\alpha & \partial\psi_{1}/\partial\lambda^1  & \partial\psi_{1}/\partial\lambda^0 \\
          \partial\psi_{0}/\partial\alpha & \partial\psi_{0}/\partial\lambda^1  & \partial\psi_{0}/\partial\lambda^0 \\
            \end{bmatrix} = \begin{bmatrix}
          \partial\psi_{\alpha}/\partial\alpha & 0_{p \times 1} & 0_{p \times 1} \\
          \partial\psi_{1}/\partial\alpha & -W_i(\alpha)I(A_i=1) & 0\\
          \partial\psi_{0}/\partial\alpha & 0 & -W_i(\alpha)I(A_i=0)\\
            \end{bmatrix}
\] where $\partial\psi_{\alpha}/\partial\alpha$ is the $p \times p$ Jacobian matrix of partial derivatives for $\psi_{\alpha}$, $\partial\psi_{a}/\partial\alpha$ are gradient vectors for $a \in \{0,1\}$, and $0_{p \times 1}$ are vectors of $0$. Then, 
\[A(\Lambda)= \begin{bmatrix} A_1 & 0_{p \times 2} \\ 
                                       A_2 & I_{2 \times 2} \\ \end{bmatrix}
                                      \] where $A_1=E(-\partial\psi_{\alpha} / \partial \alpha)$ and $A_2=[E(-\partial\psi_1 / \partial\alpha), E(-\partial\psi_0 / \partial\alpha)]^T$. Let \[B(\Lambda)=\begin{bmatrix} B_{11} & B_{21}^{T} \\
                                             B_{21} & B_{22}
\end{bmatrix}\]where $B_{11}$ is $(p \times p)$, $B_{21}$ is $2 \times p$, and $B_{22}$ is $(2 \times 2)$. By Lemma 7.3.11 in \cite{Casella}, $A_1=B_{11}$. It is straightforward to show that $A_2=B_{21}$. Thus, \[V(\Lambda)=\begin{bmatrix} A_1^{-1} & 0_{p \times 2} \\ 
                                           -A_2A_1^{-1} & I_{2 \times 2} \\ \end{bmatrix} \begin{bmatrix} A_1 & A_2^{T} \\ 
                                           A_2 & B_{22} \\ \end{bmatrix}\begin{bmatrix} A_1^{-1} &  -(A_1^{-1})^TA_2^T\\ 
                                         0_{2 \times p}   & I_{2 \times 2} \\ \end{bmatrix} 
                                         = \begin{bmatrix} A_1^{-1} &  0_{p \times 2}\\ 
                                         0_{2 \times p}   & B_{22}-A_2A_1^{-1}A_2^T\\ \end{bmatrix}\] Letting $g(\Lambda)=\lambda^1/\lambda^0$, it then follows from the delta method that 
\[\sqrt{n} \left( \frac{\hat{\lambda}^1_{IPTW}}{\hat{\lambda}^0_{IPTW}} - \frac{\lambda^1}{\lambda^0} \right) \xrightarrow[]{d} N \left(0, \Sigma_{IPTW}^* \right) \]
where
\[\Sigma_{IPTW}^*=\frac{\partial g}{\partial \Lambda} V(\Lambda) \left(\frac{\partial g}{\partial \Lambda}\right)^T=g^{*T}B_{22}g^*-g^{*T}(A_2A_1^{-1}A_2^T)g^* = \Sigma_{IPTW}-g^{*T}(A_2A_1^{-1}A_2^T)g^*
\] and $g^* = [1 / \lambda^0 \hspace{.3cm} -\lambda^1 / (\lambda^0)^2]$. 
The final equality holds because $B_{22}=V(\Lambda)$ from \ref{Prop1WtsFixed}. Note that $g^{*T}(A_2A_1^{-1}A_2^T)g^*=g^{*T}(A_2B_{11}^{-1}A_2^T)g^* \ge 0$ because $B_{11}$ is positive semi-definite. Thus, $\Sigma^*_{IPTW} \le \Sigma_{IPTW}$.

\subsection{Section 2.2.2 Derivations}
\label{sec222res}
Assume that $\hat{E}(Y_i \mid L_i,A_i=a)$ is estimated based on one of the four models described in Section 2.2.2. Define the set of estimating equations: \begin{equation} \label{EEPG} \sum_{i=1}^n \psi(Y_i,A_i,L_i; \gamma, \lambda) = \begin{bmatrix} \sum_{i=1}^n \psi_\gamma (Y_i, A_i, L_i; \gamma) \\
\sum_{i=1}^n \psi_1 (Y_i, L_i; \gamma, \lambda^1) \\
\sum_{i=1}^n \psi_0 (Y_i, L_i; \gamma, \lambda^0) 
\end{bmatrix} = 0\end{equation}
where $\psi_{\gamma}$ is the derivative of the log-likelihood function for the model with respect to the regression coefficients.

When the outcome model is correctly specified, these estimating equations are unbiased based on maximum likelihood theory, with solutions $\hat{\gamma}$. Now we define the estimating equations for the causal means. Define $\hat{\lambda}_{PG}^a=\hat{E}(Y^a) = \int \hat{E}(Y \mid L,A)d\hat{F}_L(l) = n^{-1} \sum_{i=1}^n \hat{E}(Y_i \mid L_i,A_i=a)$ where $\hat{F}_L$ is the empirical distribution function of $L$. Then, $\sum_{i=1}^n \psi_a (Y_i, A_i, L_i; \gamma, \lambda^a) = \sum_{i=1}^n \{ E(Y_i \mid L_i,A_i=a) - \lambda^a \} = 0$ for $a \in \{0, 1 \}$, where $E(Y_i \mid L_i,A_i=a)$ is estimated by the predicted count for observation $i$ based on the outcome model. When the model is correctly specified and based on causal consistency and conditional exchangeability, 
\[E\{\psi_a (Y_i, L_i; \gamma, \lambda^a)\} 
= E \{E(Y_i \mid L_i,A_i=a) - \lambda^a \} 
= E\{E(Y_i^a \mid L_i,A_i=a) \} - \lambda^a \]\[
= E\{E(Y_i^a \mid L_i) \} - \lambda^a 
= E(Y^a) - \lambda^a 
= 0 \]
Thus, (\ref{EEPG}) is an unbiased set of estimating equations, implying 
\[\sqrt{n}  \begin{bmatrix}  \hat{\gamma} - \gamma\\
          \hat{\lambda}^1_{PG} - \lambda^1\\
          \hat{\lambda}^0_{PG} - \lambda^0
            \end{bmatrix} 
  \xrightarrow[]{d} N \left(0, \Sigma_{PG} \right)
\]where $\Sigma_{PG}=A(\Lambda)^{-1} B(\Lambda) \{A(\Lambda)^{-1}\}^T$, with $\Lambda^T=(\gamma^T, \lambda^1, \lambda^0)$, $A(\Lambda)=E\{-\dot{\psi}(Y_i,A_i,L_i,\Lambda)\}$, 

\noindent{$B(\Lambda)=E\{\psi(Y_i,A_i,L_i,\Lambda)\psi(Y_i,A_i,L_i,\Lambda)^T\}$}, and $\dot{\psi}(Y_i,A_i,L_i,\Lambda)=\partial \psi(Y_i,A_i,L_i,\Lambda)/ \partial \Lambda^T$. The delta method can then be applied to obtain the asymptotic distribution of $\widehat{CMR}_{PG}=\hat{\lambda}_{PG}^1/\hat{\lambda}_{PG}^0$. Specifically, let $g(\Lambda)=\lambda^1/\lambda^0$ such that $\partial g(\Lambda) / \partial (\Lambda) =  [0_{1 \times p}, 1/\lambda^0, -\lambda^1/(\lambda^0)^2]^T$. Then,
\[\sqrt{n} \left( \frac{\hat{\lambda}_{PG}^1}{\hat{\lambda}_{PG}^0} - \frac{\lambda^1}{\lambda^0} \right) \xrightarrow[]{d} N \left(0, \Sigma_{PG}^* \right)
\]
where \[\Sigma_{PG}^*=\frac{\partial g(\Lambda)}{\partial (\Lambda)} \Sigma_{PG} \left(\frac{\partial g(\Lambda)}{\partial (\Lambda)}\right)^T\]

\subsection{Section 2.2.3 Derivations}
Define the set of estimating equations: \[ \sum_{i=1}^n \psi(Y_i,A_i,L_i; \alpha, \gamma, \lambda) = \begin{bmatrix} \sum_{i=1}^n \psi_\alpha (A_i, L_i; \alpha) \\ \sum_{i=1}^n \psi_\gamma (Y_i, A_i, L_i; \gamma) \\
\sum_{i=1}^n \psi_1 (Y_i, A_i, L_i; \alpha, \gamma, \lambda^1) \\
\sum_{i=1}^n \psi_0 (Y_i, A_i, L_i; \alpha, \gamma, \lambda^0) 
\end{bmatrix} = 0\]
where $\psi_{\alpha}$ and $\psi_{\gamma}$ are defined in Sections \ref{IPTWresults} and \ref{sec222res} and 
\begin{align}
\psi_1 (Y, A, L; \alpha, \gamma, \lambda^1) &=  [AY-\{A-e(L, \alpha)\}m_1(L,\gamma)]\{e(L, \alpha)\}^{-1} - \lambda^1 \nonumber \\ \psi_0 (Y, A, L; \alpha, \gamma, \lambda^0) &=  \{(1-A)Y+\{A-e(L, \alpha)\}m_0(L,\gamma)\}\{1-e(L, \alpha)\}^{-1} - \lambda^0 \nonumber \end{align} are the estimating equations for $\hat{\lambda}_{DR}^1$ and $\hat{\lambda}_{DR}^0$, respectively.

When either the weight model or the outcome model is correctly specified, the solutions $\hat{\lambda}^1_{DR}$ and $\hat{\lambda}^0_{DR}$ to the estimating equations
\noindent{$\psi_1 (Y_i, A_i, L_i; \alpha, \gamma, \lambda^1)$} and $\psi_0 (Y_i, A_i, L_i; \alpha, \gamma, \lambda^0)$ are consistent estimators of the causal means $\lambda^1$ and $\lambda^0$, respectively. This is shown as follows. Suppose $\hat{\alpha} \overset{p}{\to} \alpha_0$ and $\hat{\gamma} \overset{p}{\to} \gamma_0$, where $\overset{p}{\to}$ denotes convergence in probability. When the weight model is correctly specified, $e(L_i,\alpha_0)=P(A_i=1 \mid L_i)$. Similarly, when the outcome model is correctly specified, $m_a(L_i,\gamma_0)=E(Y_i \mid L_i,A_i=a)$. By causal consistency and with algebraic manipulation, $\psi_1 (Y_i, A_i, L_i; \alpha, \gamma, \lambda^1) =  Y_i^1+\{A_i-e(L_i, \alpha)\}\{Y_i^1-m_1(L_i,\gamma)\}\{e(L_i, \alpha)\}^{-1}- \lambda^1$. Note that by conditional exchangeability: \[E[\{A_i-e(L_i, \alpha)\}\{Y_i^1-m_1(L_i,\gamma)\}\{e(L_i, \alpha)\}^{-1}]\]\[=E_L \left( \{e(L_i, \alpha)\}^{-1} E_{A \mid L}\{A_i-e(L_i, \alpha)\} E_{Y^1 \mid L}\{Y_i^1-m_1(L_i,\gamma)\} \right)\] When the weight model is correctly specified, $E_{A \mid L}\{A_i-e(L_i, \alpha)\}=E_{A \mid L}\{A_i\}-{e}(L_i, \alpha_0)=0$ and when the outcome model is correctly specified $E_{Y^1 \mid L}\{Y_i^1-m_1(L_i,\gamma)\}=E_{Y^1 \mid L}\{Y_i^1\}-m_1(L_i,\gamma_0)=0$. Then, $E \{ \psi_1 (Y_i, A_i, L_i; \alpha, \gamma, \lambda^1)\} = E(Y^1) - \lambda^1 = 0$. Thus, 

\noindent{$\psi_1 (Y_i, A_i, L_i; \alpha, \gamma, \lambda^1)$} is unbiased when the weight or outcome model is correctly specified. Similarly, $\psi_0 (Y_i, A_i, L_i; \alpha, \gamma, \lambda^0)$ is unbiased when either model is correctly specified. 
Therefore, 
\[\sqrt{n} \begin{bmatrix} \hat{\alpha} - \alpha \\                  \hat{\gamma} -\gamma \\
          \hat{\lambda}^1_{DR} - \lambda^1 \\
          \hat{\lambda}^0_{DR} - \lambda^0
            \end{bmatrix} \xrightarrow[]{d} N \left(0, \Sigma_{DR} \right)
\]where $\Sigma_{DR}=A(\Lambda)^{-1} B(\Lambda) \{A(\Lambda)^{-1}\}^T$, with $\Lambda^T=(\alpha^T, \gamma^T, \lambda^1, \lambda^0)$, $A(\Lambda)=E\{-\dot{\psi}(Y_i,A_i,L_i,\Lambda)\}$, 

\noindent{$B(\Lambda)=E\{\psi(Y_i,A_i,L_i,\Lambda)\psi(Y_i,A_i,L_i,\Lambda)^T\}$}, and $\dot{\psi}(Y_i,A_i,L_i,\Lambda)=\partial \psi(Y_i,A_i,L_i,\Lambda)/ \partial \Lambda^T$. The delta method is applied to obtain the asymptotic distribution of $\widehat{CMR}_{DR}=\hat{\lambda}_{DR}^1/\hat{\lambda}_{DR}^0$. Specifically, let $g(\Lambda)=\lambda^1/\lambda^0$ such that $\partial g(\Lambda) / \partial (\Lambda) = [0_{1 \times c}, 0_{1 \times p}, 1 / \lambda^0, -\lambda^1 / (\lambda^0)^2]$, where $c$ and $p$ are the number of coefficients in the weight and outcome models, respectively. Then, 
\[\sqrt{n} \left( \frac{\hat{\lambda}_{DR}^1}{\hat{\lambda}_{DR}^0} - \frac{\lambda^1}{\lambda^0} \right) \xrightarrow[]{d} N \left(0, \Sigma_{DR}^* \right)
\]
where \[\Sigma_{DR}^*=\frac{\partial g(\Lambda)}{\partial (\Lambda)} \Sigma_{DR} \left(\frac{\partial g(\Lambda)}{\partial (\Lambda)}\right)^T\]

\subsection{Section 2.3.1: Motivation for the plug-in heaping estimators}
 
 Under the assumed heaping model (dropping subscripts $i$ for notational ease), $Y_h^a=\Delta Y^a+(1-\Delta) h_{\eta}(Y^a)$ for $a \in \{0,1\}$. Under the assumption that $\Delta \perp Y^a$, \[E(Y_h^a)=E(\Delta) E(Y^a)+E\{(1-\Delta)\} E\{h_{\eta}(Y^a)\}\] Because $E(\Delta)=\pi$, this implies \[E(Y_h^a)=\pi E(Y^a)+(1-\pi) E\{h_{\eta}(Y^a)\}\] and therefore
\[E(Y^a)=\pi^{-1} [ E(Y_h^a)-(1-\pi)E\{h_{\eta}(Y^a)\}] \]

\section*{Appendix B: Multiple Imputation}
\setcounter{subsection}{0}
\subsection*{B1 Number of Partners}
The WIHS data were reanalyzed using multiple imputation as a sensitivity analysis to the main analysis conducted in Sections 4.1 of the manuscript. Specifically, the longitudinal WIHS sample of 4,982 women was limited to the 3,378 women who were alive and attended at least one visit during the study period (2007-2017). Participant age and study site were imputed deterministically for missed visits. Data from the 21 potential visits during the study period were then formatted into a wide data set, with the values of site, age, HIV status, education, history of incarceration prior to the study period, race/ethnicity, sexual orientation, and marital status at the start of the study period treated as time fixed and incarceration status, the number of male sexual partners, alcohol and drug use, pot use, sex exchange practices, and unstable housing treated as time varying. Multiple imputation via fully conditional specification was conducted using the MI procedure in SAS Version 9.4. Thirty imputed data sets were generated from the imputation model following 10 burn-in iterations. 

The analytic sample was then derived from each imputed data set. For each woman who was incarcerated between 2007-2017, her first incarcerated visit following a non-incarcerated visit was selected as her baseline visit. Covariates from the visit preceding baseline were included in the anlaysis, and the outcome was measured at the visit following baseline. This resulted in between $n=343$ and $n=363$ (mean $n=351$) incarcerated women across the 30 multiply imputed data sets. As in the primary anlaysis, a sample of visits from women who did not report being incarcerated between 2007-2017 was randomly selected, stratifying by visit number to ensure the same distribution of baseline visits over calendar time as the incarcerated women. The resulting WIHS samples ranged from $1,029$ to $1,089$ (mean $1,054$) women across the 30 imputed data sets. Each sample was analysed using the methods described in Section 4.1, except that the NB distribution was used for the PG and DR estimators rather than the ZINB due to convergence issues with the ZINB model for some multiply imputed data sets. The 30 sets of results were combined using Rubin's method \citep{rubin2004multiple}. The estimated $CMRs$ (95\% CIs) for the IPTW, parametric g-formula, and doubly robust methods were 1.24 (0.39, 2.08), 1.29 (0.86, 1.72), and 1.35 (0.29, 2.40), respectively. These results are similar to those presented in Section 4.1 of the main text, but are less precise.

An additional sensitivity analysis was conducted where the number of sexual partners at baseline (categorized as 0, 1, or 2+) was included in the weight and/or outcome models. The estimated $CMRs$ (95\% CIs) for the IPTW, parametric g-formula, and doubly robust methods were 1.17 (0.35, 1.99), 1.23 (0.86, 1.60), and 1.27 (0.21, 2.34), respectively. These results are similar to the results that exclude this covariate.

\subsection*{B2 Number of Cigarettes}
The number of cigarettes outcome was analyzed analogously. In the multiple imputation model, the sex exchange practices, sexual orientation, and marital status covariates were removed while household income was added as a time-varying covariate. The outcome variable was replaced with the number of cigarettes smoked per day. The analytic sample was obtained as with the number of partners outcome, except that both incarcerated women and the sample of unincarcerated women were restricted to those who reported smoking at the visit prior to baseline. This resulted in between $n=233$ and $n=253$ (mean $n=242$) incarcerated women across the 30 multiply imputed data sets and between $n=932$ and $n=1012$ (mean $n=969$) total women in the sample.

As in the main analysis, the NB distribution was used for all heaping estimators. HCAR estimators and the informative heaping estimator were applied, where the probability of reporting an exact count was allowed to vary across the following heaping intervals: $<5$, $5-14$, and $15+$. The estimated $CMRs$ (95\% CIs) for the HCAR IPTW, parametric g-formula, and doubly robust methods were 1.19 (0.87, 1.52), 1.09 (0.89, 1.30), and 1.19 (0.86, 1.52), respectively. For the informative heaping estimator, the estimated $CMR$ (95\% CI) was 1.09 (0.89, 1.30). These estimates are slightly larger than the primary findings, but corresponding confidence intervals overlap. 

An additional sensitivity analysis was conducted where the number of cigarettes at baseline (categorized as $<10$ or $10+$) was included in the weight and/or outcome models. The estimated $CMRs$ (95\% CIs) for the HCAR IPTW, parametric g-formula, and doubly robust methods were 1.19 (0.87, 1.51), 1.11 (0.91, 1.31), and 1.19 (0.86, 1.53), respectively. For the informative heaping estimator, the estimated $CMR$ (95\% CI) was 1.11 (0.91, 1.31). These findings are similar to the results that exclude this covariate.

\section*{MACS/WIHS Combined Cohort Study (MWCCS) Full Acknowledgement}
The contents of this publication are solely the responsibility of the authors and do not represent the official views of the National Institutes of Health (NIH). MWCCS (Principal Investigators): Atlanta CRS (Ighovwerha Ofotokun, Anandi Sheth, and Gina Wingood), U01-HL146241; Bronx CRS (Kathryn Anastos and Anjali Sharma), U01-HL146204; Brooklyn CRS (Deborah Gustafson and Tracey Wilson), U01-HL146202; Data Analysis and Coordination Center (Gypsyamber D’Souza, Stephen Gange and Elizabeth Golub), U01-HL146193; Chicago-Cook County CRS (Mardge Cohen and Audrey French), U01-HL146245; Northern California CRS (Bradley Aouizerat, Jennifer Price, and Phyllis Tien), U01-HL146242; Metropolitan Washington CRS (Seble Kassaye and Daniel Merenstein), U01-HL146205; Miami CRS (Maria Alcaide, Margaret Fischl, and Deborah Jones), U01-HL146203; UAB-MS CRS (Mirjam-Colette Kempf, Jodie Dionne-Odom, and Deborah Konkle-Parker), U01-HL146192; UNC CRS (Adaora Adimora), U01-HL146194. The MWCCS is funded primarily by the National Heart, Lung, and Blood Institute (NHLBI), with additional co-funding from the Eunice Kennedy Shriver National Institute Of Child Health \& Human Development (NICHD), National Institute On Aging (NIA), National Institute Of Dental \& Craniofacial Research (NIDCR), National Institute Of Allergy And Infectious Diseases (NIAID), National Institute Of Neurological Disorders And Stroke (NINDS), National Institute Of Mental Health (NIMH), National Institute On Drug Abuse (NIDA), National Institute Of Nursing Research (NINR), National Cancer Institute (NCI), National Institute on Alcohol Abuse and Alcoholism (NIAAA), National Institute on Deafness and Other Communication Disorders (NIDCD), National Institute of Diabetes and Digestive and Kidney Diseases (NIDDK), National Institute on Minority Health and Health Disparities (NIMHD), and in coordination and alignment with the research priorities of the National Institutes of Health, Office of AIDS Research (OAR). MWCCS data collection is also supported by UL1-TR000004 (UCSF CTSA), UL1-TR003098 (JHU ICTR), P30-AI-050409 (Atlanta CFAR), P30-AI-073961 (Miami CFAR), P30-AI-050410 (UNC CFAR), P30-AI-027767 (UAB CFAR), and P30-MH-116867 (Miami CHARM).

\setcounter{figure}{0}    
\renewcommand\thefigure{A\arabic{figure}}    
\renewcommand{\theHfigure}{A\arabic{figure}}

\setcounter{table}{0}    
\renewcommand\thetable{A\arabic{table}}    
\renewcommand{\theHtable}{A\arabic{table}}

\begin{figure}[p]
\begin{center}
  \includegraphics[width=\linewidth]{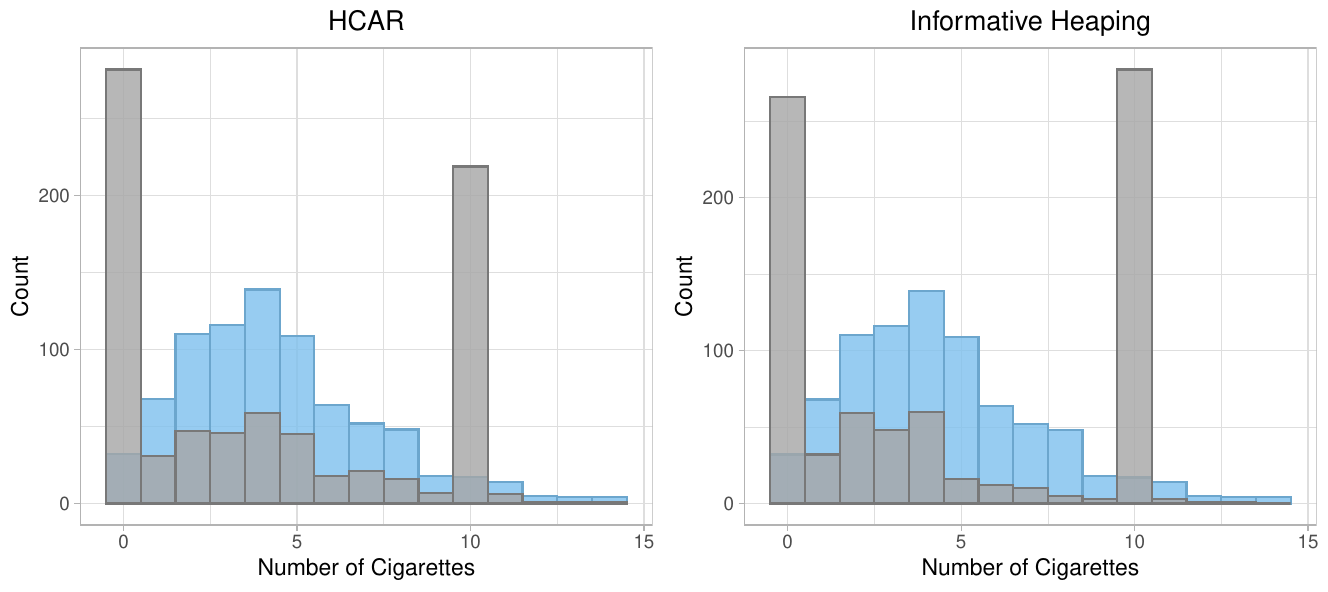}
  \caption{Histograms of the true distribution of cigarettes (blue) and the reported, i.e., heaped, number of cigarettes (gray) for a single simulated sample under Scenario 1 (HCAR) and Scenario 2 (Informative Heaping), $n=800$}
  \label{Fig:HeapSim}
\end{center}
\end{figure}

\begin{center}
\begin{table}[p]
\label{webtab1}
\centering
\caption{Results of the simulation study without data heaping by distribution and method across $5000$ samples with correct model specification, $n=800$. Percent empirical bias, MSE, ESE, SER, and empirical 95\% CI coverage calculated for the $CMR$.}
\begin{tabular}{l l c c c c c} 
\hline
  Distribution & Method & \makecell{Empirical \\ Bias (\%)} & MSE & ESE & SER & \makecell{95\% CI \\ Coverage (\%)} \\ 
 \hline
Poisson	&	IPTW, fixed	&	0.4	&	0.33	&	0.11	&	2.92	&	100	\\
	&	IPTW, estimated	&	0.4	&	0.11	&	0.11	&	0.98	&	95	\\
	&	PG	&	0.0	&	0.08	&	0.08	&	0.98	&	95	\\
	&	DR	&	0.0	&	0.08	&	0.08	&	0.98	&	95	\\
NB	&	IPTW, fixed	&	1.6	&	0.41	&	0.27	&	1.49	&	99	\\
	&	IPTW, estimated	&	1.6	&	0.26	&	0.27	&	0.94	&	94	\\
	&	PG	&	0.5	&	0.16	&	0.16	&	0.98	&	95	\\
	&	DR	&	1.3	&	0.24	&	0.26	&	0.95	&	94	\\
ZIP	&	IPTW, fixed	&	0.5	&	0.34	&	0.12	&	2.76	&	100	\\
	&	IPTW, estimated	&	0.5	&	0.12	&	0.12	&	0.96	&	95	\\
	&	PG	&	0.2	&	0.08	&	0.08	&	0.98	&	94	\\
	&	DR	&	0.2	&	0.09	&	0.09	&	0.95	&	94	\\
ZINB	&	IPTW, fixed	&	1.6	&	0.42	&	0.28	&	1.47	&	99	\\
	&	IPTW, estimated	&	1.6	&	0.27	&	0.28	&	0.94	&	94	\\
	&	PG	&	0.4	&	0.19	&	0.17	&	1.08	&	96	\\
	&	DR	&	1.4	&	0.29	&	0.27	&	1.09	&	95	\\

 \hline
\end{tabular}

\footnotesize{Abbreviations: IPTW=Inverse Probability of Treatment Weight; PG=Parametric g-formula; DR=Doubly Robust Estimator; MSE=Median Estimated Standard Error; ESE=Empirical Standard Error; SER=Standard Error Ratio (MSE/ESE); CI=Confidence Interval; CMR=Causal Mean Ratio; NB=Negative Binomial; ZIP=Zero-Inflated Poisson; ZINB=Zero-Inflated Negative Binomial. Note: ZIP PG and DR results exclude 3.5\% of simulations where models did not converge. ZINB PG and DR results exclude 2.5\% and 2.4\% of simulations, respectively, where models did not converge.}
\label{tab:P3Table1}
\end{table}
\end{center}

\begin{center}
\begin{table}[p]
\label{webtab2}
\centering
\caption{Results of the simulation study without data heaping by distribution and method across $5000$ samples with the weight model misspecified (MW), the outcome model misspecified (MO), or both models misspecified (MB), $n=800$. Percent empirical bias, MSE, ESE, SER, and empirical 95\% CI coverage calculated for the $CMR$.}
\begin{tabular}{l l c c c c c} 
\hline
  Distribution & Method & \makecell{Empirical \\ Bias (\%)} & MSE & ESE & SER & \makecell{95\% CI \\ Coverage (\%)} \\ 
 \hline
Poisson	&	IPTW, fixed, MW	&	7.2	&	0.35	&	0.13	&	2.76	&	100	\\	
	&	IPTW, estimated, MW	&	7.2	&	0.13	&	0.13	&	0.98	&	87	\\	
	&	DR, MW	&	0.0	&	0.08	&	0.08	&	0.98	&	95	\\	
NB	&	IPTW, fixed, MW	&	8.5	&	0.43	&	0.29	&	1.48	&	100	\\	
	&	IPTW, estimated, MW	&	8.5	&	0.27	&	0.29	&	0.95	&	94	\\	
	&	DR, MW	&	1.2	&	0.24	&	0.25	&	0.95	&	94	\\	
ZIP	&	IPTW, fixed, MW	&	7.5	&	0.36	&	0.14	&	2.63	&	100	\\	
	&	IPTW, estimated, MW	&	7.5	&	0.13	&	0.14	&	0.97	&	88	\\	
	&	DR, MW	&	0.2	&	0.09	&	0.09	&	0.95	&	94	\\	
ZINB	&	IPTW, fixed, MW	&	8.4	&	0.44	&	0.30	&	1.45	&	100	\\	
	&	IPTW, estimated, MW	&	8.4	&	0.28	&	0.30	&	0.94	&	94	\\	
	&	DR, MW	&	1.3	&	0.29	&	0.27	&	1.09	&	96	\\	\hline
Poisson	&	PG, MO	&	7.2	&	0.12	&	0.13	&	0.98	&	87	\\	
	&	DR, MO	&	0.4	&	0.11	&	0.12	&	0.98	&	95	\\	
NB	&	PG, MO	&	9.0	&	0.18	&	0.19	&	0.98	&	91	\\	
	&	DR, MO	&	1.6	&	0.26	&	0.27	&	0.95	&	94	\\	
ZIP	&	PG, MO	&	7.4	&	0.13	&	0.13	&	0.97	&	87	\\	
	&	DR, MO	&	0.6	&	0.12	&	0.13	&	0.96	&	94	\\	
ZINB	&	PG, MO	&	8.9	&	0.21	&	0.19	&	1.08	&	94	\\	
	&	DR, MO	&	1.6	&	0.30	&	0.28	&	1.08	&	96	\\	\hline
Poisson	&	DR, MB	&	7.2	&	0.12	&	0.13	&	0.98	&	87	\\	
NB	&	DR, MB	&	8.5	&	0.27	&	0.29	&	0.95	&	94	\\	
ZIP	&	DR, MB	&	7.4	&	0.13	&	0.14	&	0.97	&	88	\\	
ZINB	&	DR, MB	&	8.4	&	0.32	&	0.30	&	1.07	&	96	\\	

	\hline
\end{tabular}

\footnotesize{Abbreviations: IPTW=Inverse Probability of Treatment Weight; PG=Parametric g-formula; DR=Doubly Robust Estimator; MSE=Median Estimated Standard Error; ESE=Empirical Standard Error; SER=Standard Error Ratio (MSE/ESE); CI=Confidence Interval; CMR=Causal Mean Ratio; NB=Negative Binomial; ZIP=Zero-Inflated Poisson; ZINB=Zero-Inflated Negative Binomial. Note: ZIP PG and DR results exclude 0.3\%-3.5\% of simulations where models did not converge. ZINB PG and DR results exclude 1.2\%-2.4\% of simulations where models did not converge.}
\label{tab:P3Table2}
\end{table}
\end{center}

\begin{center}
\begin{table}[p]
\centering
\caption{Results of the simulation study without data heaping by distribution and method across $5000$ samples with correct model specification, $n=2000$. Percent empirical bias, MSE, ESE, SER, and empirical 95\% CI coverage calculated for the $CMR$. }
\begin{tabular}{l l c c c c c} 
\hline
  Distribution & Method & \makecell{Empirical \\ Bias (\%)} & MSE & ESE & SER & \makecell{95\% CI \\ Coverage (\%)} \\ 
  \hline
Poisson	&	MSM, fixed	&	0.1	&	0.21	&	0.07	&	2.94	&	100	\\
	&	MSM, estimated	&	0.1	&	0.07	&	0.07	&	1.00	&	95	\\
	&	PG	&	0.1	&	0.05	&	0.05	&	0.99	&	95	\\
	&	DR	&	0.1	&	0.05	&	0.05	&	0.99	&	95	\\
NB	&	MSM, fixed	&	0.8	&	0.26	&	0.17	&	1.52	&	100	\\
	&	MSM, estimated	&	0.8	&	0.17	&	0.17	&	0.98	&	95	\\
	&	PG	&	0.2	&	0.10	&	0.10	&	1.00	&	95	\\
	&	DR	&	0.7	&	0.16	&	0.16	&	0.98	&	94	\\
ZIP	&	MSM, fixed	&	0.1	&	0.21	&	0.08	&	2.80	&	100	\\
	&	MSM, estimated	&	0.1	&	0.08	&	0.08	&	0.99	&	95	\\
	&	PG	&	0.1	&	0.05	&	0.05	&	0.97	&	95	\\
	&	DR	&	0.0	&	0.06	&	0.06	&	0.97	&	95	\\
ZINB	&	MSM, fixed	&	0.6	&	0.26	&	0.18	&	1.51	&	100	\\
	&	MSM, estimated	&	0.6	&	0.17	&	0.18	&	0.98	&	94	\\
	&	PG	&	0.2	&	0.11	&	0.11	&	1.04	&	96	\\
	&	DR	&	0.5	&	0.18	&	0.17	&	1.06	&	96	\\

 \hline

 \hline
\end{tabular}

\footnotesize{Abbreviations: IPTW=Inverse Probability of Treatment Weight; PG=Parametric g-formula; DR=Doubly Robust Estimator; MSE=Median Estimated Standard Error; ESE=Empirical Standard Error; SER=Standard Error Ratio (MSE/ESE); CI=Confidence Interval; CMR=Causal Mean Ratio; NB=Negative Binomial; ZIP=Zero-Inflated Poisson; ZINB=Zero-Inflated Negative Binomial. Note: ZIP PG and DR results exclude 0.8\% of simulations where models did not converge. ZINB PG and DR results exclude 1.2\% of simulations where models did not converge. }
\label{tab:TableS1}
\end{table}
\end{center}

\begin{center}
\begin{table}[p]
\centering
\caption{Results of the simulation study without data heaping by distribution and method across $5000$ samples with the weight model misspecified (MW), the outcome model misspecified (MO), or both models misspecified (MB), $n=2000$. Percent empirical bias, MSE, ESE, SER, and empirical 95\% CI coverage calculated for the $CMR$.  }
\begin{tabular}{l l c c c c c} 
\hline
  Distribution & Method & \makecell{Empirical \\ Bias (\%)} & MSE & ESE & SER & \makecell{95\% CI \\ Coverage (\%)} \\ 
\hline
Poisson	&	MSM, fixed, MW	&	6.9	&	0.22	&	0.08	&	2.75	&	100	\\	
	&	MSM, estimated, MW	&	6.9	&	0.08	&	0.08	&	0.99	&	73	\\	
	&	DR, MW	&	0.1	&	0.05	&	0.05	&	0.99	&	95	\\	
NB	&	MSM, fixed, MW	&	7.7	&	0.27	&	0.18	&	1.50	&	100	\\	
	&	MSM, estimated, MW	&	7.7	&	0.18	&	0.18	&	0.98	&	92	\\	
	&	DR, MW	&	0.7	&	0.16	&	0.16	&	0.98	&	94	\\	
ZIP	&	MSM, fixed, MW	&	7.0	&	0.23	&	0.09	&	2.66	&	100	\\	
	&	MSM, estimated, MW	&	7.0	&	0.09	&	0.09	&	0.99	&	75	\\	
	&	DR, MW	&	0.0	&	0.06	&	0.06	&	0.97	&	95	\\	
ZINB	&	MSM, fixed, MW	&	7.5	&	0.28	&	0.19	&	1.49	&	100	\\	
	&	MSM, estimated, MW	&	7.5	&	0.18	&	0.19	&	0.98	&	92	\\	
	&	DR, MW	&	0.4	&	0.18	&	0.17	&	1.06	&	96	\\	\hline
Poisson	&	PG, MO	&	6.9	&	0.08	&	0.08	&	0.99	&	72	\\	
	&	DR, MO	&	0.1	&	0.07	&	0.07	&	1.00	&	95	\\	
NB	&	PG, MO	&	8.5	&	0.11	&	0.11	&	1.00	&	80	\\	
	&	DR, MO	&	0.8	&	0.17	&	0.17	&	0.98	&	95	\\	
ZIP	&	PG, MO	&	6.9	&	0.08	&	0.08	&	0.99	&	74	\\	
	&	DR, MO	&	0.1	&	0.08	&	0.08	&	0.99	&	95	\\	
ZINB	&	PG, MO	&	8.8	&	0.13	&	0.12	&	1.05	&	86	\\	
	&	DR, MO	&	0.6	&	0.19	&	0.18	&	1.07	&	96	\\	\hline
Poisson	&	DR, MB	&	6.9	&	0.08	&	0.08	&	0.99	&	72	\\	
NB	&	DR, MB	&	7.7	&	0.18	&	0.18	&	0.98	&	92	\\	
ZIP	&	DR, MB	&	7.0	&	0.08	&	0.09	&	0.99	&	75	\\	
ZINB	&	DR, MB	&	7.5	&	0.20	&	0.19	&	1.07	&	95	\\

\hline
\end{tabular}

\footnotesize{Abbreviations: IPTW=Inverse Probability of Treatment Weight; PG=Parametric g-formula; DR=Doubly Robust Estimator; MSE=Median Estimated Standard Error; ESE=Empirical Standard Error; SER=Standard Error Ratio (MSE/ESE); CI=Confidence Interval; CMR=Causal Mean Ratio; NB=Negative Binomial; ZIP=Zero-Inflated Poisson; ZINB=Zero-Inflated Negative Binomial. Note: ZIP PG and DR results exclude 0.0\%-0.8\% of simulations where models did not converge. ZINB PG and DR results exclude 0.4\%-1.2\% of simulations where models did not converge.}
\label{tab:TableS2}
\end{table}
\end{center}

\begin{center}
\begin{table}[p]
\centering
\caption{Results of the data heaping simulation study for Scenario 1 (HCAR) by method across $5000$ samples with correct model specification (unless otherwise noted), weight model misspecification (MW), outcome model misspecification (MO), or both models misspecified (MB), $n=800$. Percent empirical bias, MSE, ESE, SER, and empirical 95\% CI coverage calculated for the $CMR$. }
\begin{tabular}{l l c c c c c c} 
\hline
  Method & Estimator & \makecell{Empirical \\ Bias (\%)} & MSE & ESE & SER & \makecell{95\% CI \\ Coverage (\%)} \\ 
 \hline
IPTW, fixed	&	Naïve	&	6.7	&	0.10	&	0.08	&	1.20	&	93	\\
	&	HCAR	&	0.3	&	0.10	&	0.09	&	1.09	&	97	\\
	&	HCAR, MW	&	-11.6	&	0.09	&	0.09	&	0.98	&	59	\\
IPTW, estimated	&	Naïve	&	6.7	&	0.08	&	0.08	&	0.99	&	84	\\
	&	HCAR	&	0.3	&	0.09	&	0.09	&	1.00	&	95	\\
	&	HCAR, MW	&	-11.6	&	0.09	&	0.09	&	0.98	&	58	\\
PG	&	Naïve	&	7.3	&	0.08	&	0.08	&	0.99	&	82	\\
	&	HCAR	&	0.1	&	0.05	&	0.05	&	0.99	&	95	\\
	&	HCAR, MO	&	-12.6	&	0.05	&	0.06	&	0.98	&	18	\\
	&	IH	&	0.1	&	0.05	&	0.05	&	0.99	&	95	\\
	&	IH, MO	&	-12.6	&	0.05	&	0.06	&	0.98	&	18	\\
DR	&	Naïve	&	6.7	&	0.08	&	0.08	&	0.99	&	83	\\
	&	HCAR	&	0.3	&	0.09	&	0.09	&	1.00	&	95	\\
	&	HCAR, MW	&	0.0	&	0.09	&	0.09	&	0.98	&	95	\\
	&	HCAR, MO	&	-0.1	&	0.09	&	0.09	&	1.00	&	95	\\
	&	HCAR, MB	&	-11.6	&	0.09	&	0.09	&	1.00	&	63	\\
 \hline
\end{tabular}

\footnotesize{Abbreviations: HCAR=Heaping Completely at Random; IH=Informative Heaping; IPTW=Inverse Probability of Treatment Weight; PG=Parametric g-formula; DR=Doubly Robust Estimator; MSE=Median Estimated Standard Error; ESE=Empirical Standard Error; SER=Standard Error Ratio (MSE/ESE); CI=Confidence Interval; CMR=Causal Mean Ratio. Note: Results exclude one simulation where models did not converge.}
\end{table}
\end{center}

\begin{center}
\begin{table}[p]
\centering
\caption{Results of the data heaping simulation study for Scenario 1 (HCAR) by method across $5000$ samples with correct model specification (unless otherwise noted), weight model misspecification (MW), outcome model misspecification (MO), or both models misspecified (MB), $n=2000$. Percent empirical bias, MSE, ESE, SER, and empirical 95\% CI coverage calculated for the $CMR$. }
\begin{tabular}{l l c c c c c c} 
\hline
  Method & Estimator & \makecell{Empirical \\ Bias (\%)} & MSE & ESE & SER & \makecell{95\% CI \\ Coverage (\%)} \\ 
 \hline
IPTW, fixed	&	Naïve	&	6.5	&	0.06	&	0.05	&	1.20	&	78	\\
	&	HCAR	&	0.1	&	0.06	&	0.06	&	1.10	&	97	\\
	&	HCAR, MW	&	-11.6	&	0.06	&	0.06	&	1.01	&	27	\\
IPTW, estimated	&	Naïve	&	6.5	&	0.05	&	0.05	&	0.99	&	63	\\
	&	HCAR	&	0.1	&	0.06	&	0.06	&	1.01	&	95	\\
	&	HCAR, MW	&	-11.6	&	0.06	&	0.06	&	1.01	&	27	\\
PG	&	Naïve	&	7.1	&	0.05	&	0.05	&	0.99	&	57	\\
	&	HCAR	&	0.0	&	0.03	&	0.03	&	0.99	&	95	\\
	&	HCAR, MO	&	-12.8	&	0.03	&	0.03	&	0.99	&	1	\\
	&	IH	&	0.0	&	0.03	&	0.03	&	0.99	&	95	\\
	&	IH, MO	&	-12.8	&	0.03	&	0.03	&	0.99	&	1	\\
DR	&	Naïve	&	6.5	&	0.05	&	0.05	&	0.99	&	61	\\
	&	HCAR	&	0.1	&	0.06	&	0.06	&	1.01	&	95	\\
	&	HCAR, MW	&	0.1	&	0.06	&	0.06	&	1.01	&	95	\\
	&	HCAR, MO	&	0.1	&	0.06	&	0.06	&	1.03	&	96	\\
	&	HCAR, MB	&	-11.6	&	0.06	&	0.06	&	1.03	&	35	\\
 \hline
\end{tabular}

\footnotesize{Abbreviations: HCAR=Heaping Completely at Random; IH=Informative Heaping; IPTW=Inverse Probability of Treatment Weight; PG=Parametric g-formula; DR=Doubly Robust Estimator; MSE=Median Estimated Standard Error; ESE=Empirical Standard Error; SER=Standard Error Ratio (MSE/ESE); CI=Confidence Interval; CMR=Causal Mean Ratio}
\end{table}
\end{center}

\begin{center}
\begin{table}[p]
\centering
\caption{Results of the data heaping simulation study for Scenario 2 (Informative Heaping) by method across $5000$ samples with correct model specification (unless otherwise noted), weight model misspecification (MW), outcome model misspecification (MO), or both models misspecified (MB), $n=800$. Percent empirical bias, MSE, ESE, SER, and empirical 95\% CI coverage calculated for the $CMR$. }
\begin{tabular}{l l c c c c c c} 
\hline
  Method & Estimator & \makecell{Empirical \\ Bias (\%)} & MSE & ESE & SER & \makecell{95\% CI \\ Coverage (\%)} \\ 
 \hline
IPTW, fixed	&	Naïve	&	4.4	&	0.09	&	0.07	&	1.19	&	96	\\
	&	HCAR	&	-4.5	&	0.08	&	0.07	&	1.10	&	89	\\
	&	HCAR, MW	&	-13.9	&	0.07	&	0.07	&	0.99	&	30	\\
IPTW, estimated	&	Naïve	&	4.4	&	0.07	&	0.07	&	0.98	&	90	\\
	&	HCAR	&	-4.5	&	0.07	&	0.07	&	1.00	&	85	\\
	&	HCAR, MW	&	-13.9	&	0.07	&	0.07	&	0.99	&	29	\\
PG	&	Naïve	&	4.9	&	0.07	&	0.08	&	0.98	&	88	\\
	&	HCAR	&	-0.1	&	0.05	&	0.06	&	0.98	&	94	\\
	&	HCAR, MO	&	-13.0	&	0.05	&	0.06	&	0.98	&	16	\\
	&	IH	&	0.2	&	0.06	&	0.06	&	0.98	&	95	\\
	&	IH, MO	&	-13.0	&	0.05	&	0.06	&	0.98	&	16	\\
DR	&	Naïve	&	4.4	&	0.07	&	0.07	&	0.98	&	90	\\
	&	HCAR	&	-4.5	&	0.07	&	0.07	&	1.00	&	86	\\
	&	HCAR, MW	&	-4.8	&	0.07	&	0.07	&	0.99	&	84	\\
	&	HCAR, MO	&	-4.5	&	0.07	&	0.07	&	0.99	&	86	\\
	&	HCAR, MB	&	-13.9	&	0.07	&	0.07	&	1.00	&	34	\\
 \hline
\end{tabular}

\footnotesize{Abbreviations: HCAR=Heaping Completely at Random; IH=Informative Heaping; IPTW=Inverse Probability of Treatment Weight; PG=Parametric g-formula; DR=Doubly Robust Estimator; MSE=Median Estimated Standard Error; ESE=Empirical Standard Error; SER=Standard Error Ratio (MSE/ESE); CI=Confidence Interval; CMR=Causal Mean Ratio. Note: Results exclude one simulation where models did not converge.}
\end{table}
\end{center}

\begin{center}
\begin{table}[p]
\centering
\caption{Results of the data heaping simulation study for Scenario 2 (Informative Heaping) by method across $5000$ samples with correct model specification (unless otherwise noted), weight model misspecification (MW), outcome model misspecification (MO), or both models misspecified (MB), $n=2000$. Percent empirical bias, MSE, ESE, SER, and empirical 95\% CI coverage calculated for the $CMR$. }
\begin{tabular}{l l c c c c c c} 
\hline
  Method & Estimator & \makecell{Empirical \\ Bias (\%)} & MSE & ESE & SER & \makecell{95\% CI \\ Coverage (\%)} \\ 
 \hline
IPTW, fixed	&	Naïve	&	4.2	&	0.06	&	0.05	&	1.20	&	90	\\
	&	HCAR	&	-4.7	&	0.05	&	0.04	&	1.09	&	77	\\
	&	HCAR, MW	&	-14.0	&	0.04	&	0.04	&	1.00	&	3	\\
IPTW, estimated	&	Naïve	&	4.2	&	0.05	&	0.05	&	0.99	&	80	\\
	&	HCAR	&	-4.7	&	0.04	&	0.04	&	0.99	&	71	\\
	&	HCAR, MW	&	-14.0	&	0.04	&	0.04	&	1.00	&	3	\\
PG	&	Naïve	&	4.7	&	0.05	&	0.05	&	0.99	&	77	\\
	&	HCAR	&	-0.2	&	0.03	&	0.03	&	0.99	&	95	\\
	&	HCAR, MO	&	-13.0	&	0.03	&	0.03	&	1.00	&	0	\\
	&	IH	&	0.1	&	0.04	&	0.04	&	0.99	&	95	\\
	&	IH, MO	&	-13.0	&	0.03	&	0.03	&	1.00	&	0	\\
DR	&	Naïve	&	4.2	&	0.05	&	0.05	&	1.00	&	80	\\
	&	HCAR	&	-4.7	&	0.04	&	0.04	&	1.00	&	72	\\
	&	HCAR, MW	&	-5.0	&	0.04	&	0.04	&	1.01	&	69	\\
	&	HCAR, MO	&	-4.6	&	0.05	&	0.04	&	1.02	&	76	\\
	&	HCAR, MB	&	-14.0	&	0.04	&	0.04	&	1.01	&	9	\\

 \hline
\end{tabular}

\footnotesize{Abbreviations: HCAR=Heaping Completely at Random; IH=Informative Heaping; IPTW=Inverse Probability of Treatment Weight; PG=Parametric g-formula; DR=Doubly Robust Estimator; MSE=Median Estimated Standard Error; ESE=Empirical Standard Error; SER=Standard Error Ratio (MSE/ESE); CI=Confidence Interval; CMR=Causal Mean Ratio. Note: Results exclude two simulations where models did not converge.}
\end{table}
\end{center}

\begin{table}[p]
\begin{threeparttable}
\caption{Baseline characteristics of WIHS sample for the number of male sexual partners outcome by incarceration status}
\centering

\begin{tabular}{l l c c}

\hline
 &  & \makecell{No Incarceration \\ $n=588$} & \makecell{Incarceration \\ $n=294$}  \\
\hline
\\
Age	&	Median (Q1,Q3)	&	47 (40, 53)	&	44 (38, 50)	\\
	&	Mean (SD)	&	47 (10)	&	44 (8)	\\
	&	Min, Max	&	24, 80	&	25, 62	\\
Educational attainment	&	High school or more	&	399 (68\%)	&	163 (55\%)	\\
Race	&	Black	&	397 (68\%)	&	216 (73\%)	\\
	&	White	&	114 (19\%)	&	38 (13\%)	\\
	&	Other	&	77 (13\%)	&	40 (14\%)	\\
WIHS Site	&	Bronx or Brooklyn, NY	&	220 (37\%)	&	59 (20\%)	\\
	&	Washington, DC	&	96 (16\%)	&	31 (11\%)	\\
	&	Los Angeles, CA	&	63 (11\%)	&	24 (8\%)	\\
	&	San Francisco, CA	&	68 (12\%)	&	55 (19\%)	\\
	&	Chicago, IL	&	72 (12\%)	&	56 (19\%)	\\
	&	Southern Sites	&	69 (12\%)	&	69 (23\%)	\\
HIV positive	&		&	425 (72\%)	&	174 (59\%)	\\
Prior incarceration	&		&	183 (31\%)	&	215 (73\%)	\\
Unstable housing	&		&	7 (1\%)	&	21 (7\%)	\\
Sex exchange practices	&		&	4 (1\%)	&	26 (9\%)	\\
Alcohol use	&	None	&	322 (55\%)	&	129 (44\%)	\\
	&	1-7 drinks/week	&	197 (34\%)	&	83 (28\%)	\\
	&	>7 drinks/week	&	69 (12\%)	&	82 (28\%)	\\
Marijuana use	&		&	90 (15\%)	&	100 (34\%)	\\
Illicit drug use	&		&	28 (5\%)	&	111 (38\%)	\\

  \hline 
\end{tabular}
\begin{tablenotes}
      \item SD=Standard Deviation

    \end{tablenotes}
  \end{threeparttable}
\end{table}

\begin{table}[p]
\begin{threeparttable}
\caption{Baseline characteristics of WIHS sample for the cigarette outcome analysis by incarceration status}
\centering

\begin{tabular}{l l c c}

\hline
 &  & \makecell{No Incarceration \\ $n=537$} & \makecell{Incarceration \\ $n=179$}  \\
\hline
\\
Age	&	Median (Q1,Q3)	&	48 (42, 53)	&	45 (38, 51)	\\
	&	Mean (SD)	&	47 (8)	&	44 (8)	\\
	&	Min, Max	&	25, 75	&	26, 62	\\
Educational attainment	&	High school or more	&	304 (57\%)	&	98 (55\%)	\\
Race	&	Black	&	399 (74\%)	&	143 (80\%)	\\
	&	White	&	83 (15\%)	&	15 (8\%)	\\
	&	Other	&	55 (10\%)	&	21 (12\%)	\\
WIHS Site	&	Bronx or Brooklyn, NY	&	193 (36\%)	&	35 (20\%)	\\
	&	Washington, DC	&	59 (11\%)	&	19 (11\%)	\\
	&	Los Angeles, CA	&	30 (6\%)	&	11 (6\%)	\\
	&	San Francisco, CA	&	90 (17\%)	&	37 (21\%)	\\
	&	Chicago, IL	&	74 (14\%)	&	30 (17\%)	\\
	&	Southern Sites	&	91 (17\%)	&	47 (26\%)	\\
HIV positive	&		&	360 (67\%)	&	109 (61\%)	\\
Prior incarceration	&		&	276 (51\%)	&	142 (79\%)	\\
Unstable housing	&		&	17 (3\%)	&	14 (8\%)	\\
Household income	&	>\$12,000 per year	&	214 (40\%)	&	43 (24\%)	\\
Alcohol use	&	None	&	265 (49\%)	&	66 (37\%)	\\
	&	1-7 drinks/week	&	182 (34\%)	&	55 (31\%)	\\
	&	>7 drinks/week	&	90 (17\%)	&	58 (32\%)	\\
Marijuana use	&		&	164 (31\%)	&	75 (42\%)	\\
Illicit drug use	&		&	75 (14\%)	&	87 (49\%)	\\

  \hline 
\end{tabular}
\begin{tablenotes}
      \item SD=Standard Deviation

    \end{tablenotes}
  \end{threeparttable}
\end{table}

	\end{appendices}	
	
	\label{lastpage}
	
\end{document}